\documentclass[11pt]{article}
\usepackage{multicol}                       	
\usepackage{eurosym}                       	
\usepackage{ulem}                            	
\usepackage[noblocks]{authblk}		%
\usepackage{tabularx}                        	%
\usepackage{graphicx}                        	%
\usepackage{dcolumn}                        	%
\usepackage{bm}                        	%
\usepackage{amsmath}                        	%
\usepackage{amssymb}                        	%
\usepackage[squaren,Gray]{SIunits}  	%
\usepackage{float} 				
\usepackage{units}				%
\usepackage{amsmath}			%
\usepackage{amssymb}			%
\usepackage{color}				%
\usepackage[dvipsnames]{xcolor}		%
\usepackage{hyperref}			%
\usepackage{url}			%
\usepackage[left,columnwise]{lineno}	
\setpagewiselinenumbers
\modulolinenumbers[1]

\usepackage{comment}
\newcommand{\blue}{\textcolor{blue}}

\usepackage[bordercolor=white,backgroundcolor=gray!30,linecolor=black,colorinlistoftodos]{todonotes}
\newcommand\rsout{\bgroup\markoverwith{\textcolor{red}{\rule[0.5ex]{2pt}{0.4pt}}}\ULon}
\usepackage{array}
\newcolumntype{L}[1]{>{\raggedright\let\newline\\\arraybackslash\hspace{0pt}}m{#1}}
\newcolumntype{C}[1]{>{\centering\let\newline\\\arraybackslash\hspace{0pt}}m{#1}}
\newcolumntype{R}[1]{>{\raggedleft\let\newline\\\arraybackslash\hspace{0pt}}m{#1}}
\let\OLDthebibliography\thebibliography
\renewcommand\thebibliography[1]
{
	\OLDthebibliography{#1}
  	\setlength{\parskip}{0pt}
  	\setlength{\itemsep}{0pt plus 0.3ex}
}
\makeatletter
\def\@maketitle{%
  \newpage
  \null
  \vskip 0.em%
  \begin{center}%
  \let \footnote \thanks
    {\huge \bfseries \@title \par}%
    \vskip .0em%
    {\normalsize
      \lineskip 5.em%
      \begin{tabular}[t]{c}%
        \@author
      \end{tabular}\par}%
    {\normalsize \@date~-- v4.1} 
  \end{center}%
  \par
  \vskip .em}
\makeatother
\newcommand{\Bs}{LB$\nu$B}
\newcommand{\BIs}{LB$\nu$B-I}

\newcommand{\BIIs}{LB$\nu$B-II}

\newcommand{\BIIIs}{LB$\nu$B-III}

\newcommand{\N}{NOvA}
\newcommand{\AC}{Appearance Channel}
\newcommand{\ACs}{AC}

\newcommand{\DCs}{DC}
\newcommand{\lbnb}{LB$\nu$B}

\newcommand{\IMOs}{IMO}

\newcommand{\NMOs}{NMO}
\newcommand{\Tr}{$\theta _{13}$}

\newcommand{\Ts}{$\theta _{12}$}

\newcommand{\Ta}{$\theta _{23}$}

\newcommand{\dCP}{$\delta_\text{\tiny CP}$}

\newcommand{\dmN}{$\delta m^2_{21}$}
\newcommand{\Dm}{$\Delta m^2_{32}$}
\newcommand{\Dmuu}{$\Delta m^2_{\mu\mu}$}
\newcommand{\Dmee}{$\Delta m^2_{ee}$}
\newcommand{\aDm}{$|\Delta m^2_{32}|$}

\newcommand{\CS}{$\chi^2$}
\newcommand{\DCS}{$\Delta\chi^2$}
\newcommand{\SQCS}{$\sqrt{\Delta\chi^2}$}

\newcommand{\DCSB}{$\Delta \chi^2_\text{\tiny BOOST}$}
\newcommand{\APC}{\small APC, CNRS/IN2P3, CEA/IRFU, Observatoire de Paris, Sorbonne Paris Cit\'{e} University, 75205 Paris Cedex 13, France}

\newcommand{\LNCA}{\small LNCA Underground Laboratory, CNRS/IN2P3 - CEA, Chooz, France}
\newcommand{\UCI}{\small Department of Physics and Astronomy, University of California at Irvine, Irvine, California 92697, USA}
\newcommand{\IJC}{\small IJCLab,, Universit\'e Paris-Saclay, CNRS/IN2P3, 91405 Orsay, France}
\newcommand{\LON}{\small Departamento de F\'isica, Universidade Estadual de Londrina, 86051-990, Londrina -- PR, Brazil}

\newcommand{\PadINFN}{\small INFN, Sezione di Padova, via Marzolo 8, I-35131 Padova, Italy}
\newcommand{\PUC}{\small Department of Physics, 
Pontif\'icia Universidade Cat\'olica do Rio de Janeiro,
Rio de Janeiro, RJ, 22451-900, Brazil}
\newcommand{\SUBA}{\small SUBATECH, CNRS/IN2P3, Universit\'{e} de Nantes, IMT-Atlantique, 44307 Nantes, France}
\newcommand{\Sussex}{\small Department of Physics and Astronomy, University of
Sussex, Falmer, Brighton BN1 9QH, United Kingdom}
\newcommand{\SYSU}{\small Sun Yat-sen University, NO. 135 Xingang Xi Road, Guangzhou, China, 510275}
\newcommand{\Prague}{\small Institute of Particle and Nuclear Physics, Faculty of Mathematics and Physics, Charles University, V Hole\v{s}ovi\v{c}k\'{a}ch 2, 180 00 Prague 8, Czech Republic}
\begin{document}
\normalem 
%
%
\author[1,2,4]{Anatael~Cabrera\thanks{\scriptsize Contact:~\texttt{anatael@in2p3.fr}}}
\author[1,2]{Yang~Han\thanks{\scriptsize  Contact:~\texttt{hany88@mail.sysu.edu.cn}}}
\author[1]{Michel~Obolensky}
\author[2]{Fabien~Cavalier}
\author[2]{Jo\~ao~Coelho}
\author[2]{Diana~Navas-Nicol\'as}
\author[2,8]{Hiroshi~Nunokawa\thanks{\scriptsize  Contact:~\texttt{nunokawa@puc-rio.br}}}
\author[2]{Laurent~Simard}
\author[3]{Jianming~Bian}
\author[3]{Nitish~Nayak}
\author[3]{Juan~Pedro~Ochoa-Ricoux}
\author[7]{Bed\v{r}ich~Roskovec}
%
%
\author[5,*]{Pietro~Chimenti\thanks{\scriptsize Contact.:~\texttt{pietro.chimenti@uel.br}}}
\author[6,*]{Stefano~Dusini\thanks{\scriptsize Contact:~\texttt{stefano.dusini@pd.infn.it}}}
%
%
\author[9,2]{Mathieu~Bongrand}
\author[9]{Rebin~Karaparambil}
\author[9]{Victor~Lebrin}
\author[9]{Benoit~Viaud}
\author[9]{Frederic~Yermia}
\author[10]{Lily~Asquith}
\author[10]{Thiago~J.~C.~Bezerra}
\author[10]{Jeff~Hartnell}
\author[10]{Pierre~Lasorak}
\author[11]{Jiajie~Ling}
\author[11]{Jiajun~Liao}
\author[11]{Hongzhao~Yu}
%
%
\affil[1]{\APC}
\affil[2]{\IJC}
\affil[3]{\UCI}
\affil[4]{\LNCA}
\affil[5]{\LON}
\affil[6]{\PadINFN}
\affil[7]{\Prague}
\affil[8]{\PUC}
\affil[9]{\SUBA}
\affil[10]{\Sussex}
\affil[11]{\SYSU}
%
%
\title{\bf Synergies and Prospects for Early Resolution of the Neutrino Mass Ordering \\
 \vspace{0.2cm}
 \Large{\it entitled in arXiv:2008.11280v1 as} \\
 \huge Earliest Resolution to the Neutrino Mass Ordering?
 \vspace{0.2cm}}
\maketitle
\vspace{-0.5cm}

\noindent
\begin{center}
    \section*{\textsc{Abstract}}
	\parbox{0.94\textwidth}
	{\small \bf \noindent
	The measurement of neutrino Mass Ordering (MO) is a fundamental element for the understanding of leptonic flavour sector of the \emph{Standard Model of Particle Physics.} 
	Its determination relies on the precise measurement of $\Delta m^2_{31}$ and $\Delta m^2_{32}$ using either
	neutrino \emph{vacuum oscillations}, such as the ones studied by medium baseline reactor experiments,
	or 
	\emph{matter effect modified oscillations} such as those manifesting in long-baseline neutrino beams (\lbnb) or atmospheric neutrino experiments.
	Despite existing MO indication today, a fully resolved MO measurement ($\geq$5$\sigma$) is most likely to await for the next generation of neutrino experiments: JUNO, whose stand-alone sensitivity is $\sim$3$\sigma$, or \lbnb\ experiments (DUNE and Hyper-Kamiokande).
	Upcoming atmospheric neutrino experiments are also expected to provide precious information. 
	In this work, we study the possible context for the earliest full MO resolution.
	A firm resolution is possible even before 2028, exploiting mainly vacuum oscillation, upon the combination of JUNO and the current generation of \lbnb\ experiments (\N\ and T2K).
	This opportunity is possible thanks to a powerful synergy boosting the overall sensitivity where the sub-percent precision of $\Delta m^2_{32}$ by \lbnb\  experiments is found to be the leading order term for the MO earliest discovery.
	We also found that the comparison between \emph{matter} and \emph{vacuum} driven oscillation results enables unique discovery potential for physics beyond the Standard Model.
	}
\vspace{2cm}
\end{center}

\begin{multicols*}{2}

\section*{\textsc{Introduction}}
\vspace{-0.1cm}
\noindent
The discovery of the \emph{neutrino} ($\nu$) \emph{oscillations} phenomenon has completed a remarkable scientific endeavor lasting several decades changing forever our understanding of the leptonic sector's phenomenology of the \emph{standard model of elementary particles} (SM).
The new phenomenon was taken into account by introducing massive neutrinos and consequently neutrino flavour mixing and the possibility of violation of charge conjugation parity symmetry or CP-violation (CPV); e.g., review~\cite{Nunokawa:2007qh}.

Neutrino oscillations imply that the neutrino mass eigenstates (${\nu_1}$, ${\nu_2}$, ${\nu_3}$) spectrum is 
non-degenerate, so at least two neutrinos are massive.
Each mass eigenstate (${\nu_i}$; with $i$=1,2,3) can be regarded as a non-trivial mixture of the known neutrino flavour eigenstates (${\nu_e}$, ${\nu_\mu}$, ${\nu_\tau}$), linked to the three ($e$, $\mu$, $\tau$) respective charged leptons.
Since no significant experimental evidence beyond three families exists so far, the mixing is characterised by the 3$\times$3 so called {\it Pontecorvo-Maki-Nakagawa-Sakata} (PMNS)~\cite{Pontecorvo:1967fh,Maki:1962mu} matrix, assumed to be unitary, thus parameterised by three independent mixing angles (${\theta_{12}}$, ${\theta_{23}}$, ${\theta_{13}}$) and one CP phase (\dCP).
The neutrino mass spectra are indirectly known via the two measured {\it mass squared differences}, indicated as \dmN ($\equiv m_2^2-m_1^2$) and
$\Delta m^2_{32}$ ($\equiv m_3^2-m_2^2$),
respectively, related to the ${\nu_2}$/${\nu_1}$ and ${\nu_3}$/${\nu_2}$ pairs.
The neutrino absolute mass is not directly accessible via neutrino oscillations and remains unknown, despite considerable active research~\cite{PDG2020}.

As of today, the field is well established both experimentally and phenomenologically.
All relevant parameters (${\theta_{12}}$, ${\theta_{23}}$, ${\theta_{13}}$ and \dmN, \aDm) are known to the few percent precision.
The \dCP\ phase and the sign of \Dm, the so-called Mass Ordering (MO), remain unknown despite existing hints (i.e., \textless\,3$\sigma$ effects).
CPV processes arise if \dCP\ is different from 0 or $\pm\pi$, i.e., CP-conserving solutions.
The measurement of the MO has the peculiarity of having only a binary solution, either normal mass ordering (NMO), in case $\Delta m^2_{31} >0$, or inverted mass ordering (IMO) if $\Delta m^2_{31} <0$.
In order words, determining MO implies to know which is the lightest neutrino ${\nu_1}$ (or ${\nu_3}$), respective the case of \NMOs\ (\IMOs).
The positive sign of \dmN\ is known from solar neutrino data~\cite{Cleveland:1998nv,Hampel:1998xg,Abdurashitov:1999zd,Ahmad:2002jz,Fukuda:2001nj} combined with KamLAND~\cite{Eguchi:2002dm}, establishing the solar large mixing angle MSW~\cite{Mikheev:1986gs,Wolfenstein:1977ue} solution.
\vspace{-0.3cm}

\section*{\textsc{Mass Ordering Knowledge}}

This publication focuses on the global strategy to achieve the earliest and most robust MO determination scenario.
MO has rich implications not only for the terrestrial oscillation experiments, to be discussed in this paper, but also for non-oscillation experiments like search for neutrinoless double beta decay (e.g., review~\cite{Dolinski:2019nrj}) or from more broad aspects, from a fundamental theoretical (e.g., review~\cite{King:2003jb}), an astrophysical (e.g., review~\cite{Dighe:1999bi}), and cosmological (e.g., review~\cite{Hannestad:2016fog}) points of view.
Present knowledge from global data~\cite{PDG2020,Esteban:2018azc,deSalas:2020pgw,Capozzi:2018ubv} implies a few $\sigma$ hints on both MO and \dCP, where the latest results were reported at \emph{Neutrino 2020 Conference}~\cite{Ref_Nu2020}.
According to the latest NuFit5.0~\cite{NuFit5.0} global data analysis, \NMOs\ is favoured up to 2.7$\sigma$.
However, this preference remains fragile, as it will be explained later on.

Experimentally, MO can be addressed via three very different techniques (e.g.,~\cite{Blennow:2013oma} for earlier work):
a) medium baseline reactor experiment~\cite{Petcov:2001sy} (i.e., JUNO)
b) long-baseline neutrino beams (labeled here \Bs) and 
c) atmospheric neutrino based experiments.
MO determination by \Bs\ and atmospheric neutrinos relies on \emph{matter effects}~\cite{Mikheev:1986gs,Wolfenstein:1977ue} as neutrinos traverse the Earth over long enough baselines.
Since Earth is made of matter, and not of anti-matter, the effect of elastic forward scattering for electron anti-neutrinos and neutrinos depends on the sign of \Dm.
Instead, JUNO~\cite{An:2015jdp} is currently the only experiment able to resolve MO via dominant \emph{vacuum} oscillations \footnote{JUNO has a minor matter effect impact, mainly on the \dmN\ oscillation while tiny on MO sensitive \Dm\ oscillation~\cite{Li:2016txk}.}, thus holding a unique insight and capability in the MO world strategy.

The current generation of \Bs\ experiments, here called  \BIIs \footnote{The first generation \BIs\ are here considered to be K2K~\cite{Ahn:2002up}, MINOS~\cite{Adamson:2011qu} and OPERA~\cite{Agafonova:2010dc} experiments.}, are \N~\cite{Ayres:2004js} and T2K~\cite{Abe:2011ks}.
 These are to be followed up by the next generation \BIIIs\ with the DUNE~\cite{Abi:2020evt} and the Hyper-Kamiokande (HK)~\cite{Abe:2018uyc} experiments, which are expected to start taking data around 2027. 
In Korea, a possible second HK detector would enhance its MO determination sensitivity~\cite{Abe:2016ero}.
In this paper we focus mainly on the immediate impact of the \BIIs.
Nonetheless, we shall highlight the prospect contributions by \BIIIs, due to their leading order implications to the MO resolution.
Contrary to those experiments,  JUNO relies on high precision reactor neutrino spectral analysis for the extraction of MO sensitivity.

The relevant atmospheric neutrino experiments are Super-Kamiokande~\cite{Fukuda:1998mi} (SK) and IceCube~\cite{Aartsen:2014yll} (both running) as well as future specialised facilities such as INO~\cite{Kumar:2017sdq}, ORCA~\cite{Katz:2014tta} and PINGU~\cite{Aartsen:2014oha}.
The advantage of atmospheric neutrinos experiments to probe many baselines simultaneously, is partially compensated by the more considerable uncertainties in baseline and energy reconstruction and limited $\nu/\bar\nu$ separation.
The HK experiment may also offer critical MO insight via atmospheric neutrinos. 

Despite their different MO sensitivity potential and time schedules (discussed in the end), it is worth highlighting each technique's complementarity as a function of the relevant neutrino oscillation unknowns.
The MO sensitivity of atmospheric experiments depends heavily on the so called $\theta_{23}$ \emph{octant ambiguity} \footnote{This implies the approximate degeneracy of oscillation probabilities for the cases between $\theta_{23}$ and $(\pi/4-\theta_{23})$.}~\cite{Fogli:1996pv}, while \Bs\ experiments exhibit a smaller dependence.
JUNO is, however, independent, a unique asset.
Regarding the unknown \dCP, its role in atmospheric and \Bs's inverts, while JUNO remains uniquely independent.
This way, the MO sensitivity dependence on \dCP\ is less important for 
atmospheric neutrinos (i.e. washed out), but \BIIs\ are to a great extent handicapped by the degenerate phase-space competition to resolve both \dCP\ and MO simultaneously. 
In brief, the MO sensitivity interval of ORCA/PINGU swings about the 3$\sigma$ to 5$\sigma$, depending on the value of \Ta\, and \BIIs\ sensitivities are effectively blinded to MO for more than half of the \dCP\ phase-space.
However, DUNE has the unique ability to resolve MO, also via matter effects, regardless of \dCP.
Although not playing an explicit role, the constraint on \Tr, from reactor experiments (i.e. Daya Bay~\cite{Adey:2018zwh}, 
Double Chooz~\cite{DoubleChooz:2019qbj} and RENO~\cite{Bak:2018ydk}), is critical for the MO (and \dCP) quest for JUNO and \Bs\ experiments.

This publication aims to illustrate, and numerically demonstrate, via a simplified estimation, the relevant ingredients to reach a fully resolved (i.e., $\geq$5$\sigma$) MO measurement strategy relying, whenever possible, only on existing (or imminently so) experiments to yield the fastest timeline \footnote{The timelines of experiments are involved, as the construction schedules may delay beyond the scientific teams' control. Our approach aims to provide minimal timing information to contextualise the experiments, but variations may be expected}.
Our approach relies on the latest 3$\nu$ global data information~\cite{NuFit5.0}, summarised in Table~\ref{table:mixing-parameters-nufit5.0}, to tune our analysis to the most probable and up to date measurements on \Ta, \dCP\ and \Dm, using only the \Bs\ inputs, as motivated later.
This work updates and expands previous works ~\cite{Nunokawa:2005nx,Li:2013zyd,Blennow:2013vta} basing the calculations on \Dm, instead of \Dmuu, as well as including the effects of the uncertainties on the relevant oscillation parameters. 
In addition, the here presented results are contextualized in the current experimental landscape, in terms of current precision of the oscillation parameters and the present-day performances of current and near future neutrino oscillation experiments, providing an important insight into the prospects for solving the neutrino mass ordering.

\begin{table}[H]
\begin{center}%
\vglue -0.2cm
\begin{tabular}{cccc} 
\hline 
\hline
{\bf NuFit5.0}& $\delta m^2_{21}$ & $\sin^2 \theta_{12}$ & $\sin^2 \theta_{13}$  \\
\hline 
Both MO  & 7.42$\times  10^{-5}$ eV$^2$  & 0.304 & 0.0224 \\
\hline \hline
LB$\nu$B& $\Delta m^2_{32}$ & $\sin^2 \theta_{23}$& $\delta_{\text \tiny CP}$\\
\hline 
NMO & 2.411$\times 10^{-3}$eV$^2$  & 0.565 & $-$0.91$\pi$ \\
IMO & -2.455$\times 10^{-3}$eV$^2$  & 0.568 & $-$0.46$\pi$ \\
\hline \hline
\end{tabular}%
\end{center}
\vglue -0.6cm
\caption{\small {
	In this work, the neutrino oscillation parameters are reduced to the latest values obtained in the NuFit5.0~\cite{NuFit5.0}, where \Dm, $\sin^2 \theta_{23}$ and $\delta_\text{\tiny CP}$ (last two rows) were obtained by using only LB$\nu$B experiments by fixing $\delta m^2_{21}$, $\sin^2 \theta_{12}$ and  $\sin^2 \theta_{13}$ to the values shown in this table (second row).}
	}
	\label{table:mixing-parameters-nufit5.0}
\vglue -0.3cm
\end{table}%

We also aim to highlight some important redundancies across experiments that could aid the robustness of the MO resolution and exploit -- likely for the first time -- the MO measurements for high precision scrutiny of the standard 3$\nu$ flavour scheme.
In this context, MO exploration might open the potential for manifestations of physics beyond the Standard Model (BSM), e.g., see reviews~\cite{An:2015jdp,Bandyopadhyay:2007kx}.
Our simplified approach is expected to be improvable by more complete developments (i.e. full combination of experiments' data), once data is available.
Such approach, though, is considered beyond our scope as it is unlikely to significantly change our findings and conclusions, given the data precision available today.
To better accommodate our approach's known limitations, we have intentionally performed a conservative rationale.
We shall elaborate on these points further during the discussion of the final results.
\vspace{-0.3cm}

\section*{\textsc{Mass Ordering Resolution Analysis}}

Our analysis relies on a simplified combination of experiments able to yield MO sensitivity intrinsically (i.e. standalone) and via inter-experiment synergies, where the gain may be direct or indirect. 
The indirect gain implies that the sensitivity improvement occurs due to the combination itself; i.e. hence not accessible to neither experiment alone but caused by the complementary nature of the different experiments' observables.
These effects will be carefully studied, including the delicate arising dependencies to ensure accurate prediction are obtained.
The existing synergies found embody a framework for powerful sensitivity boosting to yield MO resolution upon combination.
To this end, we shall combine the running \BIIs\ experiments with the shortly forthcoming JUNO.
The valuable additional information from atmospheric experiments will be considered qualitatively, for simplicity, only at the end during the discussion of results.
Unless otherwise stated explicitly, throughout this work, we shall use only the NuFit5.0~\cite{NuFit5.0} best-fit values summarised in  Table~\ref{table:mixing-parameters-nufit5.0}, to guide our estimations and predictions by today's data.

\begin{figure*}
	\centering
	\vglue -0.2cm
	\hglue -0.2cm
	\includegraphics[scale=0.25]{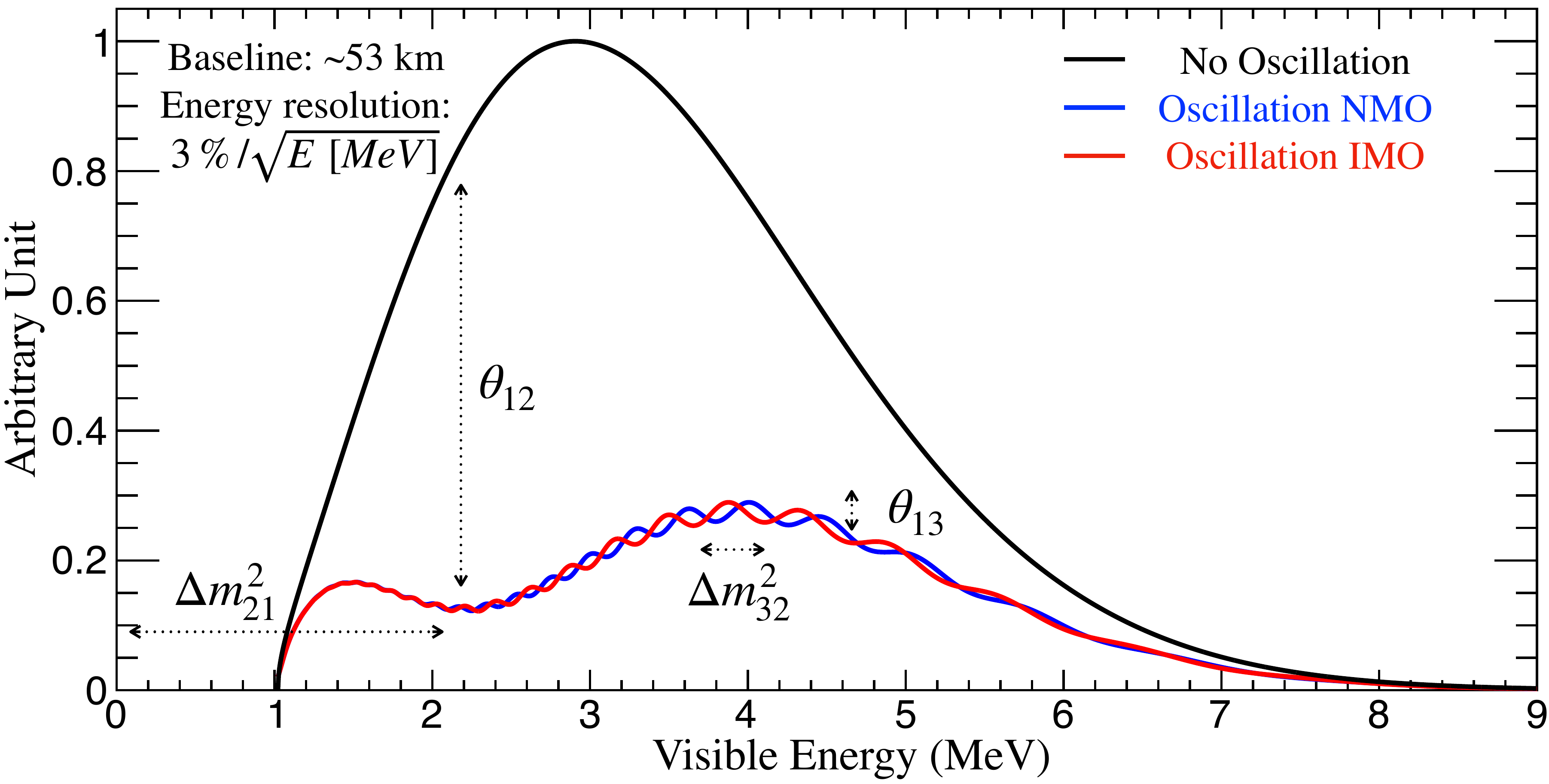}
	\vglue -0.3cm
	\caption
	{\small 
	{\bf JUNO Neutrino Bi-Oscillation Spectral Distorsion.}
	JUNO was designed to exploit the spectral distortions from two oscillations simultaneously manifesting via reactor neutrinos in a baseline of $\sim$53\,km.
	\Ts\ and \dmN\ drive the slow and large amplitude ($\sin^2 ( 2\theta_{12})/2 \approx $ 42\%) disappearance oscillation with a minimum at $\sim$2\,MeV visible energy.
	The fast and smaller amplitude  ($\sin^2 (2\theta_{13})/2 \approx $ 5\%) disappearance oscillation is driven by \Tr\ and \Dm\ instead.
	The \Tr\ oscillation frequency pattern  depends on \Dm 's sign, thus directly sensitive to mass ordering (MO) via only \emph{vacuum} oscillations.
	JUNO's high statistics allow shape-driven neutrino oscillation parameter extraction, with minimal impact from rate-only systematics.
	Hence, high precision is possible without permanent reactor flux monitoring, often referred to as \emph{near detector}(s).
	JUNO's shape analysis relies on the reactor reference spectrum's excellent control, implying high resolution, energy scale control, and a robust data-driven reference spectrum obtained with TAO~\cite{Abusleme:2020bzt}, a satellite experiment of JUNO. 
	The here presented plot 
	is for illustration purposes and the neutrino oscillation parameters are taken from NuFit5.0 (Table~\ref{table:mixing-parameters-nufit5.0}).
	}
	\label{Fig1}
	\vglue -0.3cm
\end{figure*}

\subsection*{Mass Ordering Resolution Power in JUNO}

The JUNO experiment~\cite{An:2015jdp} is one of the most powerful neutrino oscillation high precision machines. 
The JUNO spectral distortion effects are described in Figure~\ref{Fig1}, and its data-taking is expected to start in 2023~\cite{JUNO:2021vlw}.
The possibility to explore precision neutrino oscillation physics with an intermediate baseline reactor neutrino experiment was first pointed out in~\cite{Choubey:2003qx}. 
Indeed JUNO alone can yield the most precise measurements of \Ts, \dmN\ , and \aDm, at the sub-percent precision~\cite{JUNO:2021vlw} for the first time.
Therefore, JUNO will lead the precision of about half of neutrino oscillation parameters.

\begin{figure*}[t]
    \vspace{-1.cm}
	\centering
	\vglue -1.6cm
	\hglue -0.3cm
	
	\includegraphics[scale=0.6]{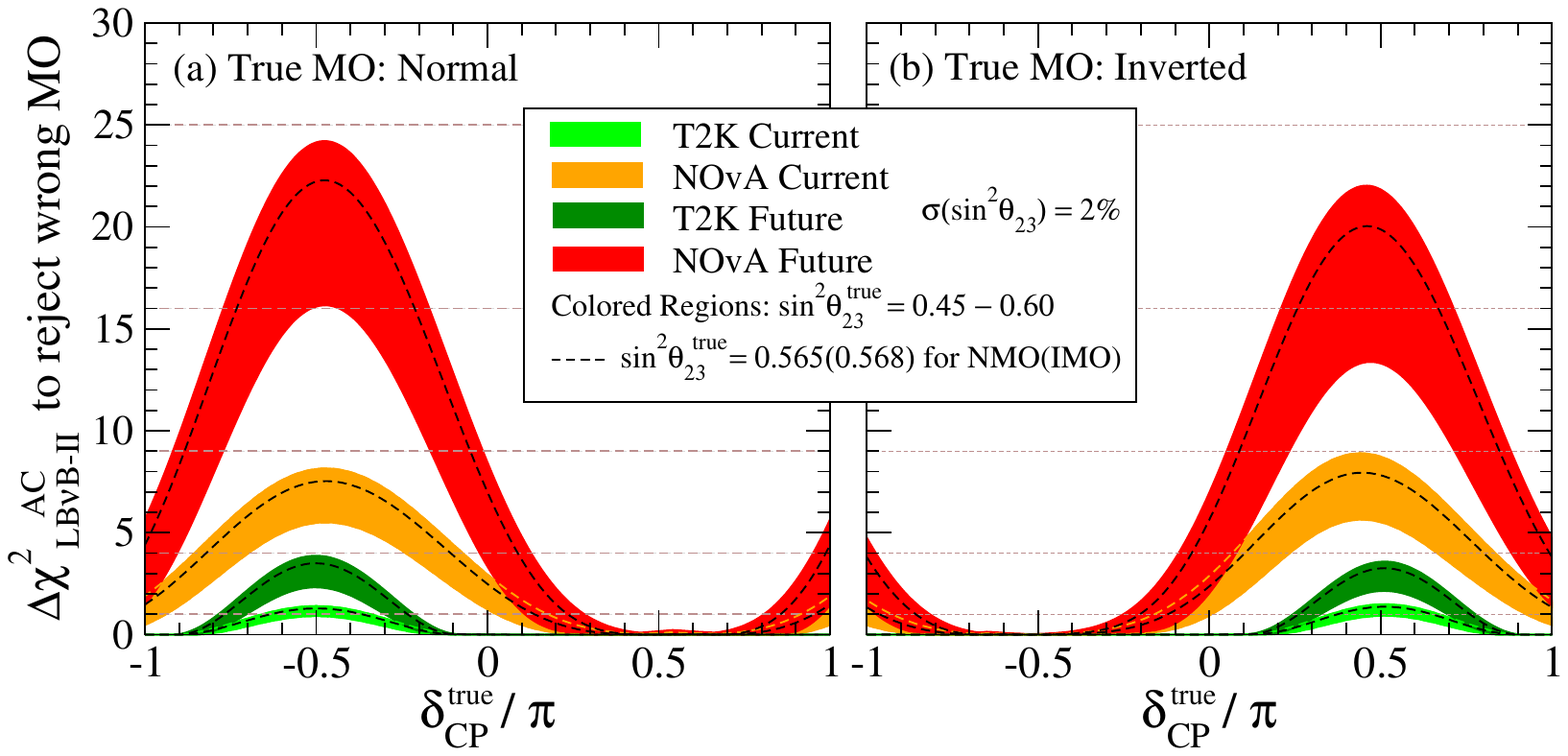}
	\vglue -1.8cm
	\vspace{-1.cm}
	\caption
	{ \small 
	{\bf \BIIs\ Mass Ordering Sensitivity.}
	The Mass Ordering (MO) sensitivity of \BIIs\ experiments via the appearance channel (\ACs), constrained to a range of $\theta_{23}$, is shown as a function of the "true" value of \dCP.
	The bands represent the cases where the "true" value of $\sin^2\theta_{23}$ lies within the interval [0.45, 0.60] with a relative experimental uncertainty of 2\%. 
	The $\sin^2\theta_{23}$ = 0.60 (0.45) gives the maximum (minimum) sensitivity for a given value of \dCP.
	The black dashed curves indicate the NuFit5.0 best fitted $\sin^2\theta_{23}$ value.
	The \NMOs\ and \IMOs\ sensitivities are illustrated respectively in the (a) and (b) panels.
	The sensitivity arises from the fake CPV effect due to matter effects, proportional to the baseline ($L$).
	The strong dependence on \dCP\ is due to the unavoidable degeneracy between NMO and IMO, thus causing the sensitivity to swig by 100\%.
	T2K, now (light green) and future (dark green), exhibits minimal intrinsic sensitivity due to its shorter baseline ($L_\text{\tiny T2K}$= 295 \,km).
	Instead, \N, now (orange) and future (red), hold leading order MO information due to its larger baseline ($L_\text{\tiny \N}$= 810\,km).
	The future full exposure for T2K and \N\ implies $\sim$3 times more statistics relative to today.
	These curves are referred to as $\Delta {\chi^2}_\text{LB$\nu$B}^\text{ AC}$ and were derived from data as detailed in Appendix~A.
	}
	\label{Fig2}
	\vglue -0.2cm
\end{figure*}

However, JUNO has been designed to yield a unique MO sensitivity via vacuum oscillation upon the spectral distortion 3$\nu$ analysis formulated in terms of \dmN\ and \Dm\ (or $\Delta m^2_{31}$).
JUNO's MO sensitivity relies on a challenging experimental articulation for the accurate control of the spectral shape-related systematics arising from energy resolution, energy scale control (nonlinearities being the most important), and even the reactor reference spectra to be measured independently by the TAO experiment~\cite{Abusleme:2020bzt}.
The nominal intrinsic MO sensitivity is $\sim$3$\sigma$ (\DCS\ $\approx$ 9) upon 6\,years of data taking.
All JUNO inputs to this paper follow the JUNO collaboration prescription~\cite{An:2015jdp}, including \Dm.
Hence, JUNO alone is unable to resolve MO with high level of confidence (\DCS $\geq$25) in a reasonable time.
In our simplified approach, we shall characterise JUNO by a simple \DCS = 9$\pm$1. 
The uncertainty aims to illustrate possible minor variations in the final sensitivity due to the experimental challenges behind or improvements in the analysis.

\vspace{-0.3cm}
\subsection*{Mass Ordering Resolution Power in \BIIs}

In all \Bs\ experiments, the intrinsic MO sensitivity arises via the \emph{appearance channel} (\ACs), from the transitions $\nu_\mu \to \nu_e$ and $\bar\nu_\mu \to \bar\nu_e$; also sensitive to \dCP.
MO manifests as an effective \emph{fake} CPV effect or bias.
This effect causes the oscillation probabilities to be different for neutrino and anti-neutrinos even under CP-conserving solutions.
It is not trivial to disentangle the genuine (\dCP) and the faked CPV terms.
Two main strategies exist, based on the fake component, which is to be either 
a) minimised (i.e. shorter baseline, like T2K, 295~km) enabling to measure mainly \dCP\ 
or 
b) maximised (i.e. longer baseline), so that matter effects are strong enough to disentangle them from the \dCP, and both can be measured simultaneously exploiting spectral information from the second oscillation maximum.
The latter implies baselines \textgreater$1000\,$km, best represented by DUNE ($1300\,$km).
\N's baseline ($810\,$km) remains a little too short for a full disentangling ability.
Still, \N\ remains the most important \Bs\ to date with sizeable intrinsic MO sensitivity due to its relatively large matter effects as compared to T2K.

Figure~\ref{Fig2} shows the current and future intrinsic MO sensitivities of \BIIs\ experiments, including their explicit \Ta\ and \dCP\ dependencies.
The obtained MO sensitivities were computed using a simplified strategy where the \ACs\ was treated as \emph{rate-only} (i.e., one-bin counting) analysis, thus neglecting any shape-driven sensitivity gain.
This approximation is remarkably accurate for off-axis beams (narrow spectrum), especially in the low statistics limit, where the impact of systematics remains small (here neglected).
The background subtraction was accounted for and tuned to the latest experiments' data.
To corroborate our estimate's accuracy, we reproduced the \BIIs\ latest results~\cite{Ref_Nu2020}, as detailed in~\textbf{Appendix~A}. 

While \N\ \ACs\ holds significant intrinsic MO information, it is unlikely to resolve (\DCS $\geq$25) alone.
This outcome is similar to that of JUNO.
Of course, the natural question may be whether their combination could yield the full resolution.
Unfortunately, as it will be shown, this is unlikely but not far.
Therefore, in the following, we shall consider their combined potential, along with T2K, to provide the extra missing push.
This may be somewhat counter-intuitive since T2K has just been shown to hold minimal intrinsic MO sensitivity, i.e., $\leq$4 units of \DCS.
Indeed, T2K, once combined, has an alternative path to enhance the overall sensitivity, which is to be described next.

\vspace{-0.3cm}

\begin{figure*}[t!]
    \vspace{-1.8cm}
    \hspace*{1cm} 
	\centering
	\includegraphics[scale=0.6]{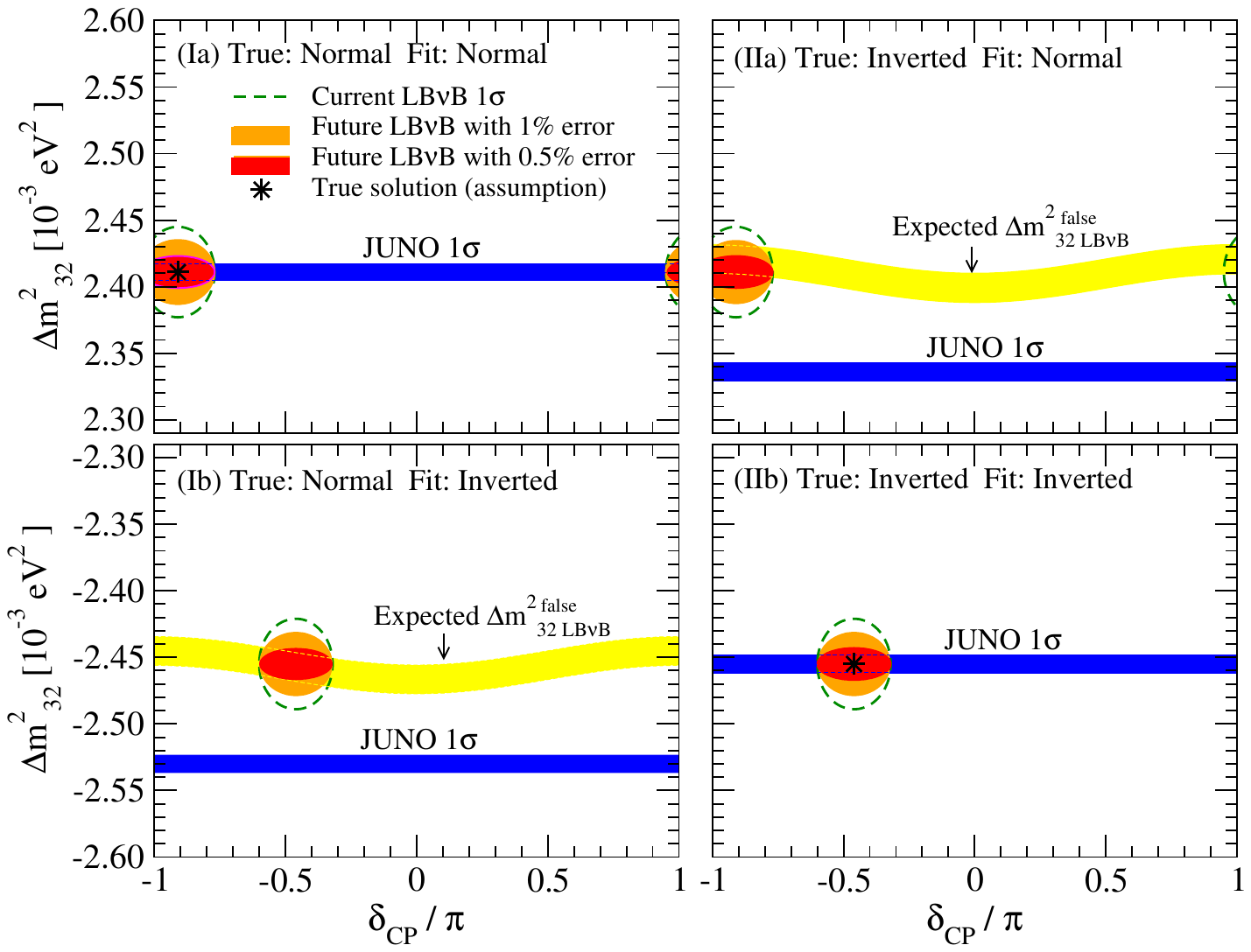}
	\vspace{-0.4cm}
	\caption{\small 
	{\bf Origin of MO Boosting by LB$\nu$B for JUNO.}
	Semi-quantitative and schematic illustration of the \Bs-JUNO MO resolution synergy is shown for the cases where the true MO is normal (left panels) or inverted (right panels). 
	For each case, the true values of $\Delta {m^2_{32}}$ are assumed to coincide with the NuFit5.0 best fitted values indicated by the black asterisk symbols.
        	For each assumed true value of $\Delta {m^2_{32}}$, possible range of the false values of $\Delta {m^2_{32}}$ to be determined from \Bs\ DC is indicated by the yellow color bands where their width reflects the ambiguity due to the CP phase (see Appendix C).
        	The approximate current 1$\sigma$ allowed ranges of ($\delta_\text{\tiny CP},\,\Delta {m^2_{32}}$) from NuFit5.0 are indicated by the dashed green curve whereas the future projections assuming the current central values with 1\% (0.5\%) uncertainty of $\Delta m^2_{32}$ are indicated by filled orange (red) color.
        	Expected 1$\sigma$ ranges of $\Delta m^2_{32}$ from JUNO alone are indicated by the blue color bands though the ones in the wrong MO region would be disfavored at $\sim 3\sigma$ confidence level (CL) by JUNO itself.
        When the MO which is assumed in the fit coincides with the true one,  allowed region of $\Delta {m^2_{32}}$ by LB$\nu$B overlaps with the one to be determined by JUNO as shown in the panels I(a) and II(b).
        	On the other hand,  when the assumed (true) MO and fitted one do not coincide, the expected (false) values of $\Delta {m^2_{32}}$ by LB$\nu$B and JUNO do not agree, as shown in the panels I(b) and II(a), disfavouring these cases, which is the origin of what we call the boosting effect in this paper.
	}
	\label{Fig3}
\end{figure*}

\subsection*{Synergetic Mass Ordering Resolution Power}

A remarkable synergy exists between JUNO and \Bs\ experiments thanks to their complementarity~\cite{An:2015jdp,Nunokawa:2005nx,Li:2013zyd,Blennow:2013vta,Minakata:2006gq,Forero:2021lax}.
In this case, we shall explore the contribution via the \Bs's {\em disappearance channel} (\DCs), i.e., the transitions $\nu_\mu \to \nu_\mu$ and $\bar\nu_\mu \to \bar\nu_\mu$.
This might appear counter-intuitive, since \DCs\ is practically blinded (i.e. variations \textless1\%) to MO, as shown in \textbf{Appendix-B}.

Instead, the \Bs\ \DCs\ provides a precise complementary measurement of \Dm.
This information unlocks a mechanism, described below, enabling the intrinsic MO sensitivity of JUNO to be enhanced by the external \Dm\ information.
This highly non-trivial synergy may yield a MO leading order role but introduces new dependences, also explored below.

Both JUNO and \Bs\ analyse data in the 3$\nu$ framework to directly provide \Dm\ (or $\Delta m^2_{31}$) as output.
The 2$\nu$ approximation leads to effective observables, such as \Dmuu\ and \Dmee~\cite{Nunokawa:2005nx} detailed in \textbf{Appendix-C}.
A CP-driven ambiguity limits the \Bs\ \DCs\ information precision on the \Dm\ measurement if \Bs\ \ACs\ measurements are not taken into account.
The role of this ambiguity is small, but not entirely negligible and will be detailed below. 
The dominant \BIIs's precision is today $\sim$2.9\% per experiment~\cite{Ref_T2K@Nu2020,NOvA:2021nfi}.
The combined \BIIs\ global precision on 
\Dm\ is already $\sim$1.4\%~\cite{NuFit5.0}.
Further improvement below 1.0\% appears possible within the \BIIs\ era when integrating the full luminosities~\cite{T2K:2016siu,NOvA:2021nfi}.
An average precision of $\sim$0.5\% is reachable only upon the next \BIIIs\ generation.
Instead, JUNO precision on \Dm\ is expected to be well within the sub-percent (\textless0.5\%) level~\cite{An:2015jdp,1975878}.

The essence of the synergy is described here.
Upon 3$\nu$ analysis, both JUNO and \Bs\ experiments obtain two different values for \Dm\ depending on the assumed MO.
Since there is only one \emph{true} solution, NMO, or IMO, the other solution is thus \emph{false}.
The standalone ability to distinguish between those two solutions is the \emph{intrinsic} MO resolution power of each experiment.
The critical observation is that the general relation between the true-false solutions is different for reactors and \Bs\ experiments, as {\it semi-quantitatively} illustrated in Figure~\ref{Fig3}.
For a given true \Dm, its false value, referred to as $\Delta {m^2_{32}}^\text{false}$, as detailed in~{\bf Appendix~C}.
This implies that both JUNO and \Bs\ based experiments generally have 2 solutions corresponding to NMO and IMO, illustrated in Figure~\ref{Fig3} by the region delimited by the dashed green ellipses for the current \Bs\, data and blue bands for JUNO.
The yellow bands indicate the possible range of false $\Delta m^2_{32}$ values expected from \Bs, including a \dCP\ dependence, if the current best fit $\Delta m^2_{32}$ is turned out to be true.

All experiments must agree on the unique true \Dm\ solution.
Consequently, the corresponding JUNO ($\Delta {m^2_{32}}^\text{false}_\text{\tiny\ JUNO}$) and \Bs\ ($\Delta {m^2_{32}}^\text{false}_\text{\tiny\ LB$\nu$B}$) false solutions will differ if the overall \Dm\ precision allows their relative resolution.
The ability to distinguish (or separate) the false solutions, or {\it mismatch} of 2 false solutions, seen in the panels (Ib) and (IIa) in
Figure~\ref{Fig3}, can be exploited as an extra dedicated discriminator expressed by the term:
\begin{eqnarray}
\Delta \chi^2_\text{\tiny BOOST} \sim 
\left(\frac{
\Delta {m^2_{32}}^\text{\tiny false}_\text{\tiny\ JUNO}-
\Delta {m^2_{32}}^\text{\tiny false}_\text{\tiny\ LB$\nu$B}}
{\sigma (\Delta m^2_{32})_\text{\tiny LB$\nu$B}}
\right)^2.
\label{Eq1}
\end{eqnarray}

This \DCSB\ term characterises the rejection of the false solutions (either NMO or IMO) through an hyperbolic dependence on the overall \Dm\ precision.
The derived MO sensitivity enhancement may be so substantial that it can be regarded and as a potential \emph{boost} effect in the MO sensitivity.

\begin{figure*}[t!]
    \vspace{-0.5cm}
	\centering
	\centering
	\includegraphics[scale=0.62]{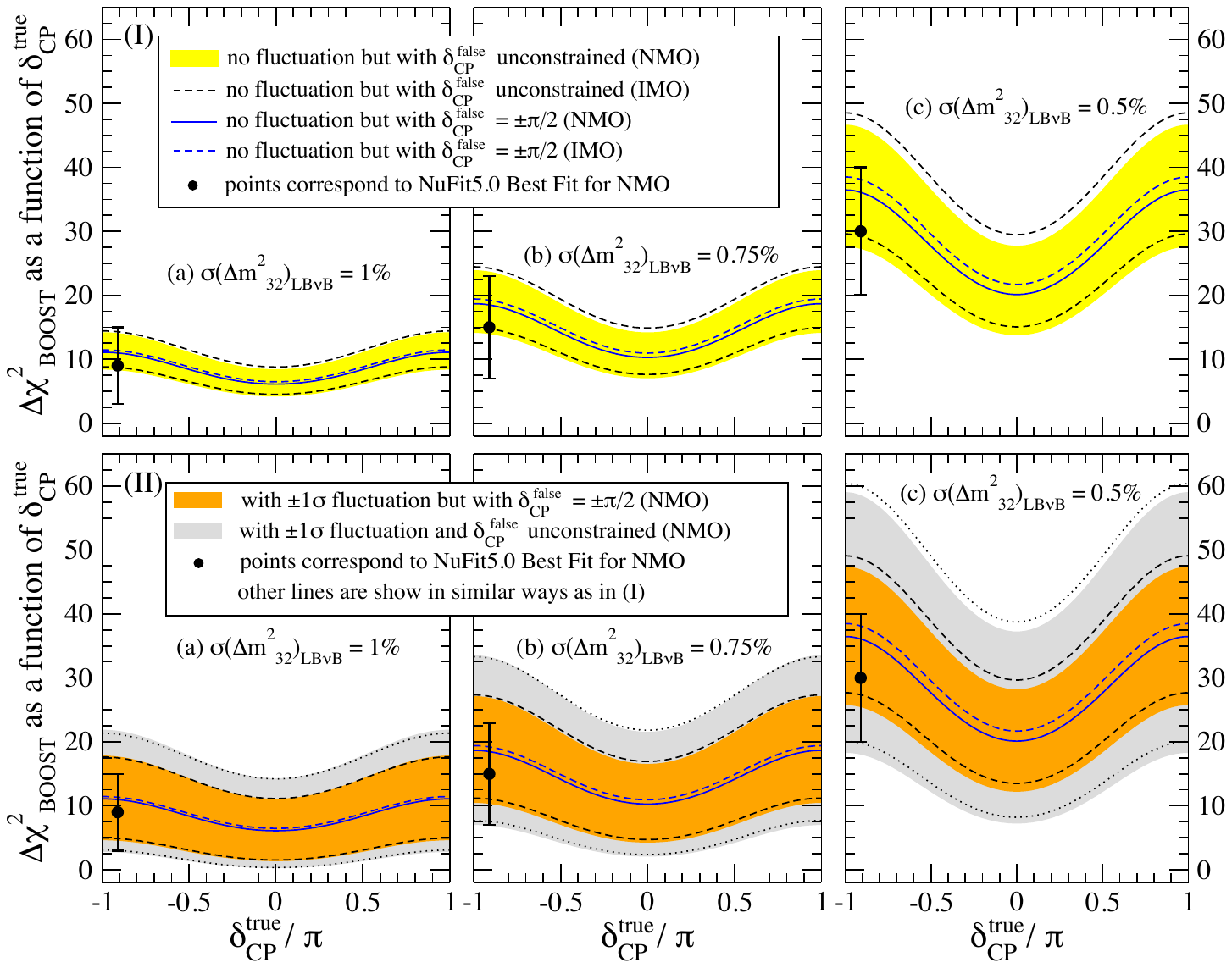}
 	\vglue -0.6cm
	\caption
	{ \small 
	{\bf JUNO and \Bs\ Mass Ordering Synergy Dependences.}
	The isolated synergy boosting term obtained from the combining JUNO and LB$\nu$B experiments is represented by 
	\DCSB, as approximately shown in Eq.~(\ref{Eq1}), see \textbf{Appendix-C} for details.
	\DCSB\ depends on the true value of $\delta_\text{\tiny CP}$ and \Dm\ precision, where uncertainties are considered:
	1.0\% (a), 
	0.75\% (b)
	and
	0.5\% (c). 
	The \DCSB\ term is almost identical for both NMO and IMO solutions.
	Two specific effects lead the uncertainty in the a priori prediction on \DCSB.
	(I) illustrates only the ambiguity of the CP phase (yellow band) impact 
	whereas
	(II) shows only the impact of the $\pm 1\sigma$ fluctuations of \Dm, as measured by \Bs\ (orange band).
	The JUNO uncertainty on \Dm\ is considered to be less than $0.5\%$.
	The grey bands {\blue in (II)} show when both effects are taken into account simultaneously.
	The mean value of the \DCSB\ term increases strongly with the precision on \Dm.
	The uncertainties from CP phase ambiguity and fluctuation could deteriorate much of the a priori gain on the prospected sensitivities.
	\Dm\ fluctuations dominate, while the \dCP\ ambiguity is only noticeable for the best \Dm\ precision.
	The use of NuFit5.0 data (black point) eliminates the impact of the \dCP\ prediction ambiguity while the impact of \Dm\ remains as fluctuations cannot be predicted a priori. 
	Today's favoured \dCP\ maximises the sensitivity gain via the \DCSB\ term.
	When quoting sensitivities, we shall consider the lowest bound as the most conservative case.
	}
	\label{Fig4}
\end{figure*}

\begin{figure*}
	\centering
	\hglue -0.3cm
	\includegraphics[scale=0.24]{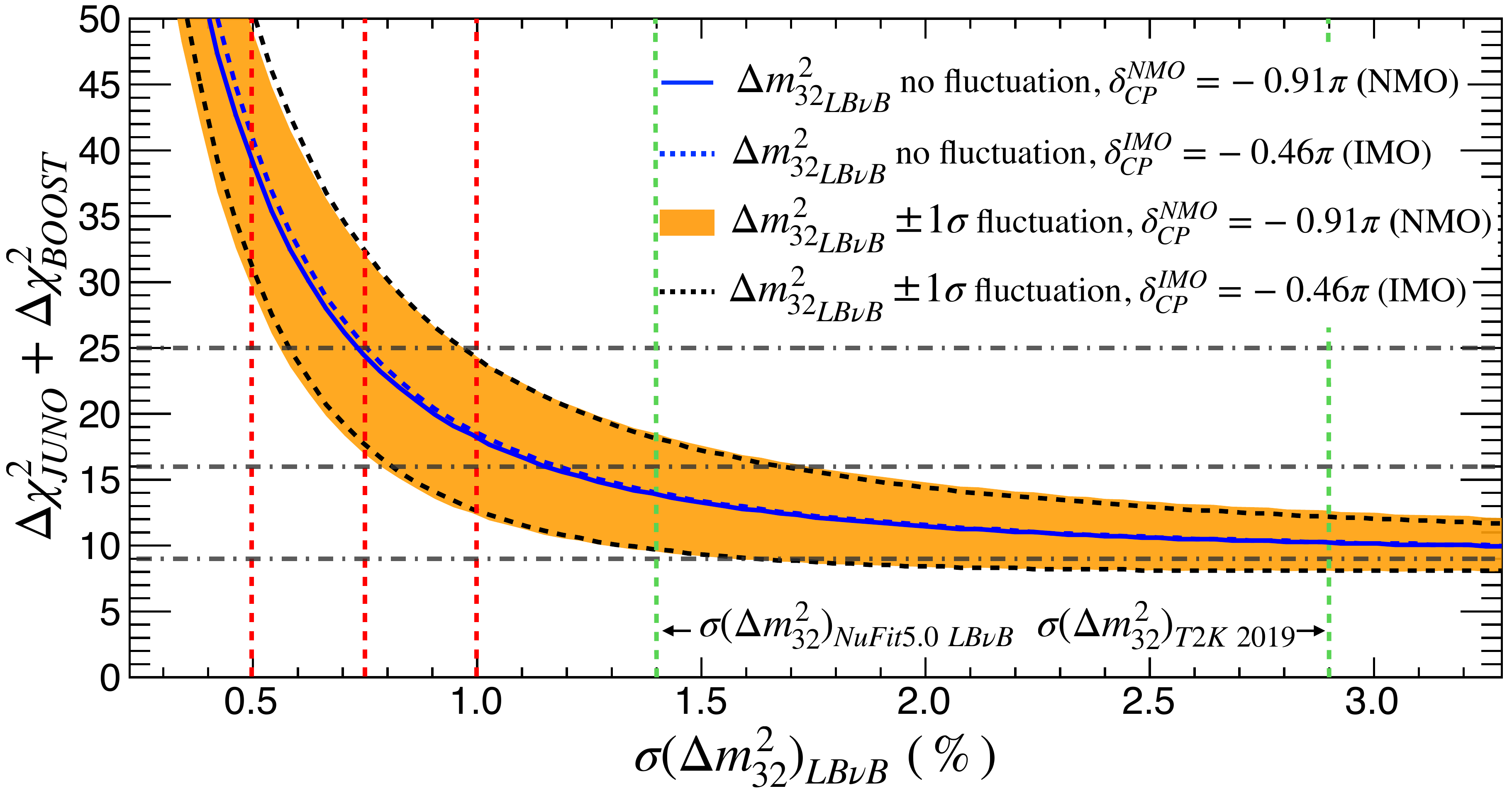}
	\caption
	{ \small 
	{\bf JUNO Mass Ordering Sensitivity Boosting.}
	A significant increase of JUNO intrinsic sensitivity ($\Delta
 \chi^2_\text{\tiny JUNO} \approx9$) is possible exploiting the \Bs's disappearance (\DCs) characterised by \DCSB\ depending strongly on the uncertainty of \Dm.
	Today's NuFit5.0 average \BIIs's precision on \Dm\ is $\sim$1.4\%. 
	A rather humble 1.0\% precision is possible, consistent with doubling the statistics if systematics allowed.
	Since \N\ and T2K are expected to increase their exposures by about factors of $\sim $3 before the shutdown, sub-percent precision may also be within reach.
	While the ultimate precision is unknown, we shall consider a $\geq$0.75\% precision to illustrate this possibility.
	So, JUNO alone (intrinsic + boosting) could yield a $\geq$4$\sigma$ (i.e., \DCS $\geq$16) MO sensitivity, at $\geq$84\% probability, within the \BIIs\ era.
	A 5$\sigma$ potential may not be impossible, depending on fluctuations.
	Similarly, JUNO may further increase in significance to resolve ($\geq$5$\sigma$ or \DCS $\geq$25) a pure vacuum oscillations MO measurement in combination with the \BIIIs's \Dm\ information.
	}
	\label{Fig5}
\end{figure*}

The JUNO-\Bs\ boosting synergy exhibits four main features as illustrated in Figure~\ref{Fig4}:

\noindent
$\bullet$ {\bf Major Increase (Boost) Potential of the Combined MO Sensitivity.} 
This is realised by the new pull term, shown in Eq.~\eqref{Eq1} and illustrated in Figure~\ref{Fig4}, which is to be added to the intrinsic MO discrimination \DCS\ terms per experiment as it will be described later on in Figures~\ref{Fig5}-\ref{Fig7}.

\noindent
$\bullet$ {\bf Dependence on the Precision of \Dm.}
Again, this is described explicitly in Eq.~(\ref{Eq1}).
The leading order effect is the uncertainty on \Dm. 
This typically referred to as $\sigma(\Delta m^2_{32})_\text{LB$\nu$B}$ as this largely dominates due to its poorer precision as compared to that obtained by JUNO ($\le$ 0.5\%) even within about a year of data-taking.
Three cases are explored in this work, 
(a) 1.0\% (i.e. close to today's precision), 
(b) 0.75\% 
and 
(c) 0.5\% (ultimate precision). 
Figure~\ref{Fig4} exhibits a strong dependence, telling us the importance of reducing the uncertainties of $\Delta m^2_{32}$ from \Bs\ to increase the MO sensitivity.
This is why T2K can have an active and important role to improve the overall MO sensitivity. 

\noindent
$\bullet$ {\bf Impact of Fluctuations.}
In order to be accurately predictive, it is important to evaluate the impact of the unavoidable fluctuations due to the today's data uncertainties on \Dm\ as well as on the \dCP\ ambiguity (see below description).
All these effects are quantified and explained in Figure~\ref{Fig4} by the orange bands, thus representing the $\pm$1$\sigma$ data fluctuations of \Dm\ from \Bs\ can significantly impact the boosted MO sensitivity.

\noindent
$\bullet$ {\bf \dCP\ Ambiguity Dependence.}
The main consequence is to limit the predictability of \DCSB, even if the assumed true value of the CP phase is fixed or limited to very narrow range.
Its effect is less negligible as the \Bs\ precision on \Dm\ improves ($\leq$0.5\%), as shown by the yellow bands in (I) and by the gray band in (II) of Fig~\ref{Fig4}.
However, by considering the $\Delta m^2_{32}$ determined by the global fit like NuFit5.0, we can reduce this ambiguity as the best fitted $\Delta m^2_{32}$ values for NMO and IMO also reflect the most likely values of $\delta_\text{\tiny CP}$ maximising our predictions' accuracy to the most probable parameter-space, as favoured by the latest world neutrino
data \footnote{Despite that $\Delta \chi^2_\text{boost}$
defined by Eqs.~\eqref{eq:chi2_juno_plus_pull} and \eqref{eq:Delta_chi2_boost}
in \textbf{Appendix-C} does not depend explicitly on the CP phase, 
we are implicitly using the CP phase information since the 
best fitted $\Delta m^2_{32}$ coming from the global analysis
carry the informtion on $\delta_{\text{\tiny CP}}$ through
the \Bs AC data used in the global analysis.}.

In brief, when combining JUNO and the \Bs\ experiments, the overall sensitivity works as if JUNO's intrinsic sensitivity gets boosted, via the external \Dm\ information.
This is further illustrated and quantified in Figure~\ref{Fig5}, as a function of the precision on \Dm\ despite the sizeable impact of fluctuations.
The \Bs\ intrinsic \ACs\ contribution will be added and shown in the next section.
It is also demonstrated that the \DCs\ information of the \Bs’s, via the boosting, play a significant role in the overall MO sensitivity.
However, this improvement cannot manifest without JUNO -- and vice versa.
For an average precision on \Dm\ below 1.0\%, even with fluctuations, the boosting effect can be already considerable.
A \Dm\ precision as good as \textgreater0.75\% may be accessible by \BIIs\ while the \BIIIs\ generation is expected to go up to $\leq$0.5\% level.

Since the exploited \DCs\ information is practically blinded to matter effects \footnote{The \Dm\ measurement depends slightly on \dCP, obtained via the \ACs\ information, itself sensitive to matter effects.}, the boosting synergy effect remains dominated by JUNO's vacuum oscillations nature.
For this reason, the sensitivity performance is almost identical for both NMO and IMO solutions, in contrast to the sensitivities obtained from solely matter effects, as shown in Figure~\ref{Fig2}.
This effect is especially noticeable in the case of atmospheric data.
The case of T2K is particularly illustrative, as its impact on MO resolution is essentially only via the boosting term mainly, given its small intrinsic MO information obtained by \ACs\ data.
This combined MO sensitivity boost between JUNO and \Bs\ (or atmospherics) is likely one of the most elegant and powerful examples so far seen in neutrino oscillations, and it is expected to play a significant role for JUNO to yield a leading impact on the MO quest, as described next.
In fact, the JUNO collaboration has already considered this effect when claiming its possible median MO sensitivity to be 4$\sigma$ potential~\cite{An:2015jdp,Li:2013zyd}.
However, JUNO prediction does not account for the \Dm\ fluctuations.
This work adds the impact of \Dm\ fluctuations and \dCP\ ambiguity on the MO discovery potential of JUNO upon boosting.
Our results are however consistent if used the same assumptions, as described in \textbf{Appendix~D}.
\vspace{-0.3cm}

\section*{Simplified Combination Rationale}

The combined MO sensitive of JUNO together with \BIIs\ experiments (\N\ and T2K) can be obtained from the independent additive of each \DCS.
Two contributions are expected: 
a) the \BIIs's \ACs, referred to as \DCS(\Bs-\ACs)
and 
b) the combined JUNO and \BIIs's \DCs, referred to as \DCS(JUNO$\oplus$\Bs-\DCs).
All terms were described in the previous sections [we use in this work the terminologies, AC (appearance channel) and DC (disappearance channel) for simplicity. This does not mean that the relevant information is coming only from AC or DC, but that $\Delta$\CS(\Bs-\ACs) comes dominantly from \Bs\ AC whereas $\Delta$\CS(JUNO$\oplus$\Bs-\DCs) comes dominantly from JUNO + \Bs\ DC].
Hence the combination can be represented as \DCS\,= \DCS(JUNO$\oplus$\Bs-\DCs) + \DCS(\Bs-\ACs), illustrated in Figure~\ref{Fig6}, where the orange and grey bands represent, respectively, the effects of the \Dm\ fluctuations and the CP-phase ambiguity. 
Figure~\ref{Fig6} quantifies the MO sensitivity in terms of significance (i.e., numbers of $\sigma$'s) obtained as \SQCS\ quantified in all previous plots.
Again, both NMO and IMO solutions are considered for 3 different cases for the \Bs\ uncertainty on \Dm.

\begin{description}

\item[The \DCS(\BIIs-\ACs) Term:]
	this is the intrinsic MO combined information, largely dominated by \N's \ACs, as described in Figure~\ref{Fig2}.
	The impact of T2K ($\leq$2$\sigma$) is minimal, but on the verge of resolving MO for the first time, T2K may still help here.
	As expected, this \DCS\ depends on \Ta\ and strongly on \dCP. 
	This is shown in Figure~\ref{Fig6} by the light green band.
        	We note that when T2K and \N\ are combined, there is $\sim 2\sigma$ significance enhancement in the positive (negative) range of $\delta_\text{\tiny CP}$ for NMO (IMO) which is not naively expected from Figure~\ref{Fig2}.
          This extra gain of sensitivity for the T2K and \N\ combined case comes from the difference of the matter effects on these experiments, and can be seen, e.g., in Figure 21 of Ref.~\cite{Abe:2014tzr}.
	The complexities of possible correlations and systematics handling of a hypothetical \N\ and T2K combination are disregarded in our study, but they are integrated within the combination of the \BIIs\ term, now obtained from NuFit5.0.
	The full \N\ data is expected to be available by 2024~\cite{Ref_NOVA@Nu2020}, while T2K will run until 2026~\cite{Ref_T2K@Nu2020}, upon the beam upgrades (T2K-II) aiming for HK.

\item[The \DCS(JUNO$\oplus$\Bs-\DCs) Term:] 
	this term can be regarded itself as composed of two contributions.
	The first part is the JUNO intrinsic information, i.e., \DCS\ = $9\pm1$ units after 6\,years of data-taking. 
	This contribution is independent of \Ta\ and \dCP, as shown in Figure~\ref{Fig6}, represented by the blue band. 
	The second part is the JUNO boosting term, shown explicitly in Figure~\ref{Fig4}, including its generic dependencies, such as the true value of \dCP.
	This term exhibits strong modulation with \dCP\ and uncertainty of \Dm, as illustrated in Figures~\ref{Fig4} and \ref{Fig5}.
	The \DCS(JUNO$\oplus$\Bs-\DCs) term strongly shapes the combined \DCS\ curves (orange).
	Indeed, this term causes the leading variation across Figure~\ref{Fig6} for the different cases of the uncertainty of \Dm:
	a) 1.0\% (top), reachable by \BIIs~\cite{T2K:2016siu,NOvA:2021nfi}, 
	b) 0.75\% (middle), maybe reachable (i.e. optimistic) by \BIIs\
	and 
	c) 0.5\% (bottom), which is only reachable by the \BIIIs\ generation~\cite{Abe:2018uyc,Abi:2020evt}.

\end{description}

\noindent
The combination of the JUNO, \ACs, and \DCs\ inputs from \BIIs\ experiments appears on the verge of achieving the first MO resolved measurement with a sizeable probability.
The combination's ultimate significance is likely to mainly depend on the final uncertainty on \Dm\ obtained by \Bs\ experiments.
The discussion of the results and implications, including limitations, is addressed in the next section.

\section*{Implications \& Discussion}

\begin{figure*}
	\centering
	\vglue -0.9cm
    \centering
	\includegraphics[scale=0.65]{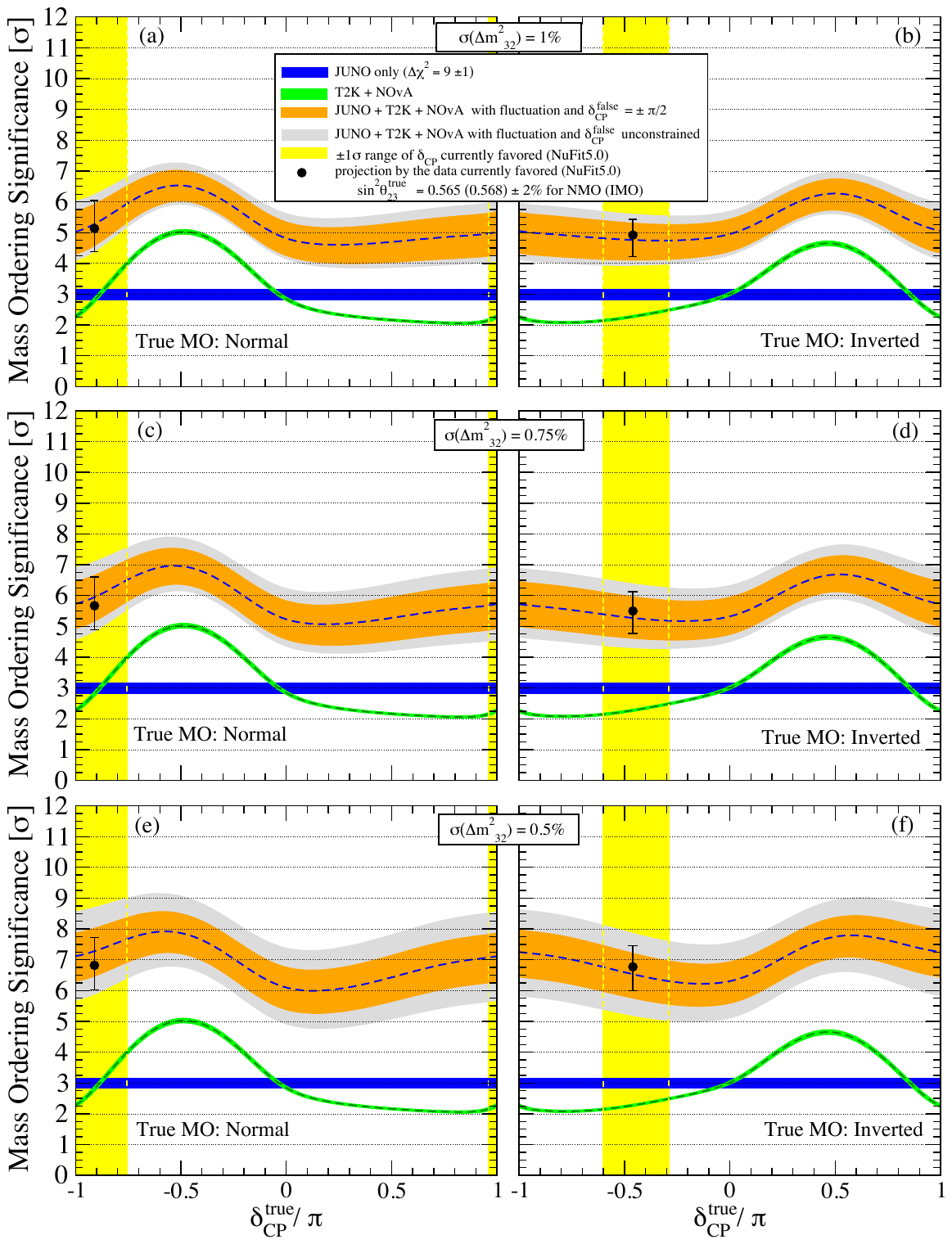}
	\vglue -1.2cm
	\caption
	{ \small 
	{\bf The Combined Mass Ordering Sensitivity.}
	The combination of the MO sensitive of JUNO and \BIIs\ is illustrated for six difference configurations: 
	NMO (left), 
	IMO (right) 
	considering the \Bs\ uncertainty on \Dm\ to 
	1.0\% (top), 
	0.75\% (middle) 
	and 
	0.5\% (bottom).
	The NuFit5.0 favoured value is set for $\sin^2 \theta_{23}$ with an assumed 2\%  experimental uncertainty.
	The intrinsic MO sensitivities are shown for JUNO (blue) and the combined \BIIs\ (green), the latter largely dominated by \N.
	The JUNO sensitivity boosts when exploiting the \Bs's \Dm\ additional information via the \DCSB\ term, described in Figure~\ref{Fig4} but not shown here for illustration simplicity. 
	The orange and grey bands illustrate the presence of the boosting term prediction effects,
        respectively, the $\pm 1\sigma$ fluctuation of \Dm\ and the \dCP\ ambiguity in addition.
	T2K impacts mainly via the precision of \Dm\ and the measurement of \dCP.
	The combined sensitivity suggests a mean (dashed blue line) $\geq$4$\sigma$ significance for any value of \dCP\ even for the most conservative $\sigma(\Delta m^2_{32})=$1\%.
	However, a robust $\geq$5.0$\sigma$ significance at 84\% probability (i.e. including fluctuations) seems possible, if the currently preferred value of
        \dCP\ and NMO remain favoured by data, as indicated by the yellow band and black point (best fit).
	Further improvement in the precision of \Dm\ translates into a better MO resolution potential.
	}
	\label{Fig6}
	\vglue -0.5cm
\end{figure*}

Possible implications arising from the main results summarised in Figure~\ref{Fig6} deserved some extra elaboration and discussion for a more accurate contextualisation, including a possible timeline and highlight the limitations associated with our simplified approach.
These are the main considerations:

\noindent
{\bf 1.~MO Global Data Trend:}
today's reasonably high significance, not far from the level to be reached by intrinsic sensitivities of JUNO or \N, is obtained by the most recent global analysis~\cite{NuFit5.0} which favours \NMOs\ up to 2.7$\sigma$.
However, this significance lowers to 1.6$\sigma$ without SK atmospherics data, thus proving their crucial value to the global MO knowledge today.
The remaining aggregated sensitivity integrates over all other experiments.
However, the global data preference is somewhat fragile, still varying between \NMOs\ and \IMOs\ solutions~\cite{Esteban:2018azc,NuFit5.0,Kelly:2020fkv}.

The reason behind this is actually the corroborating manifestation of the alluded complementarity between \BIIs\ and \emph{reactors} \footnote{Before JUNO, the reactor experiments stand for Daya Bay, Double Chooz, and RENO, whose lower precision on \Dm\ is $\sim$2\%.} experiments.
Indeed, while the current \Bs\ data alone favour IMO, the match in \Dm\ measurements by \Bs\ and reactors tend to favour the case of NMO, which is this overall solution obtained upon combination.
Hence, the MO solution currently flips due to the reactor-\Bs\ data interplay, despite the sizeable \Dm\ uncertainty fluctuations as compared to the aforementioned scenario where JUNO will be on, indicating it's crucial contribution.
This effect, expected since~\cite{Nunokawa:2005nx}, is at the heart of the described boosting mechanism and has started manifesting earlier on.
This can be regarded as the first data-driven manifestation of the aforementioned \DCSB\ effect.

\noindent{\bf 2.~Atmospherics Extra Information:}
we did not account for atmospheric neutrino input, such as the running SK and IceCube experiments.
They are expected to add valuable \DCS though susceptible to the aforementioned \Ta\ (mainly) and \dCP\ dependences.
This contribution is more complex to replicate with accuracy due to the vast $E/L$ phase-space; hence we disregarded it in our simplified analysis.
Its importance has long been proved by SK dominance of much of today's MO information.
So, all our conclusions can only be enhanced by adding the missing atmospheric contribution. 
Future ORCA and PINGU have the potential to yield extra MO information~\cite{Blennow:2013vta}, while their combinations with JUNO data is actively studied~\cite{Bezerra:2019dao,Ref_JUNO+ORCA@Nu2020} to yield full MO resolution.

\noindent{\bf 3.~Inter-Experiment Full Combination:}
a complete strategy of data-driven combination between JUNO and \BIIs\ experiments will be beneficial in the future \footnote{During the final readiness of our work, one such a combination was reported~\cite{Cao:2020ans}
using a different treatment (excluding fluctuations). While their qualitative conclusions are consistent with our studies, there may still be numerical differences left to be understood.}.
Ideally, this may be an official inter-collaboration effort to carefully scrutinise the possible impact of systematics and correlations, involving both experimental and theoretical physicists in such studies (see e.g.~\cite{Forero:2021lax}).
We do not foresee a significant change in our findings by a more complex study, including the highlighted MO discovery potential due to today's data and knowledge limitations.

Our approach did not merely demonstrate the numerical yield of the combination between JUNO and \Bs, but our goal was also to illustrate and characterise the different synergies manifesting therein.
Our study focuses on the breakdown of all the relevant contributions in the specific and isolated cases of the MO sensitivity combination of the leading experiments.
The impact of the \DCSB\ was isolated, while its effect is otherwise transparently accounted for by any complete 3$\nu$ \CS\ formulation, such as done by NuFit5.0 or other similar analyses.
Last, our study was tuned to the latest data to maximise the accuracy of predictability, which is expected to be order $\sim$0.5$\sigma$ around the 5$\sigma$ range.

\begin{figure*}
	\vspace{0.2cm}
	\centering
	\vglue -0.3cm
	\hglue -0.2cm
        	\includegraphics[scale=0.31]{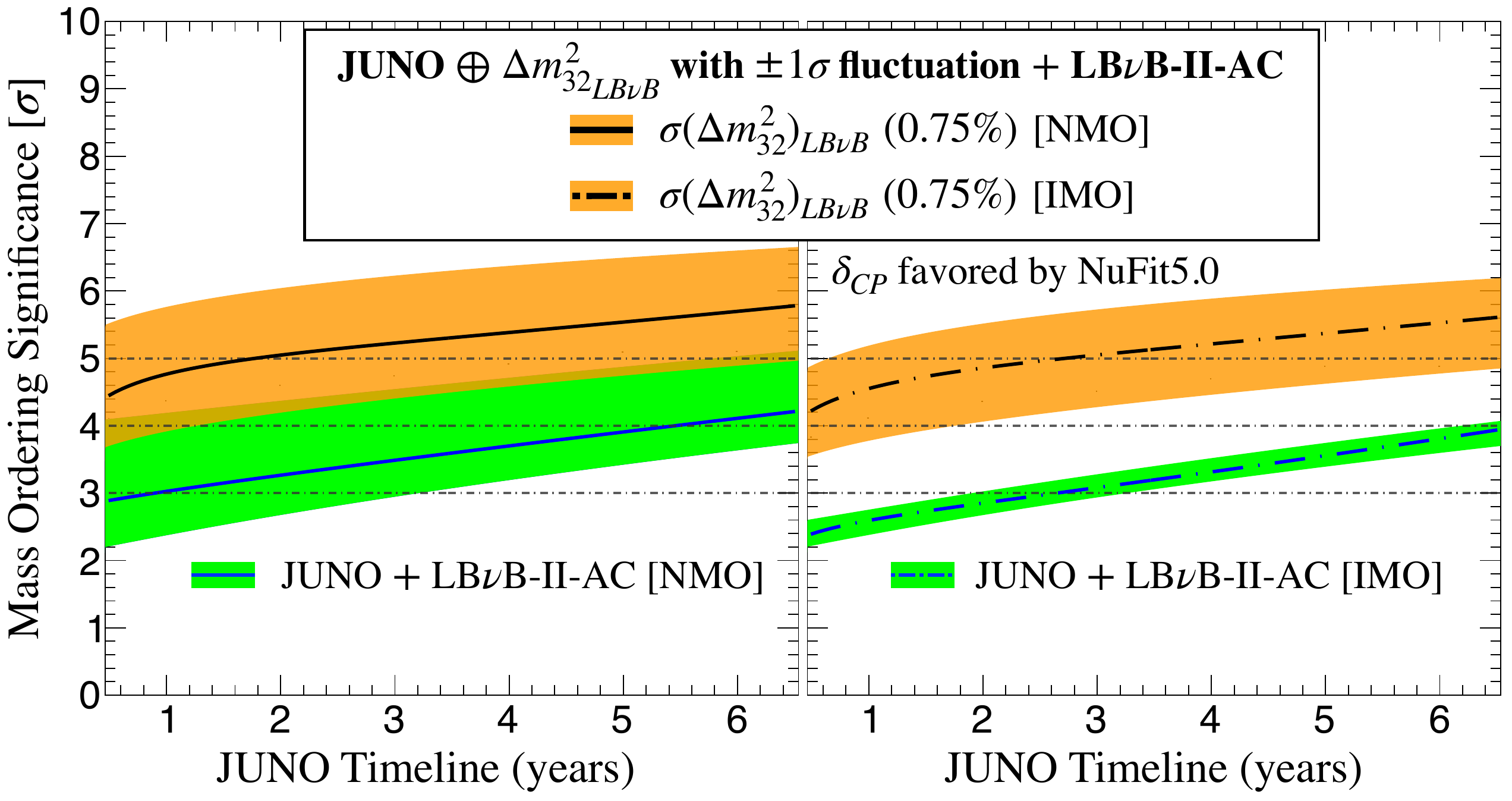}
	\vglue -0.3cm
	\caption
	{ \small 
	{\bf Mass Ordering Sensitivity \& Possible Resolution Timeline.}
	Since all the \N\ and T2K data are expected to be accumulated by 
	$\sim$2024~\cite{Ref_NOVA@Nu2020} 
	and 
	$\sim$2026~\cite{Ref_T2K@Nu2020}, the combined sensitivity follows JUNO data availability.
	JUNO is expected to start in 2023, reaching its statistically dominated nominal MO sensitivity (9~units of \DCS) within $\sim$6\,years.
	We illustrate the \NMOs\ (plot on the left) and \IMOs\ (plot on the right) scenarios.
	The sensitivity evolution depends mainly on JUNO once boosted, where 0.75\% $\Delta m_{32}^2$ uncertainty (black line) is considered.
	The effect of \Dm\ fluctuations is indicated (orange bands), including that of the variance due to the data favoured region for \dCP\ (green band).
	The larger band of the \NMOs\ is caused by contribution of the \BIIs\ experiments, whose contribution is rather negligible for opposite \IMOs.
	Since JUNO boosted dominates, the sensitivities are almost independent of \NMOs\ and \IMOs\ solutions (left), this also demonstrating the humble overall impact of the \ACs\ channel (\N\ mainly) of the \BIIs\ experiments upon combination.
	The mean significance is expected to reach the $\sim$5$\sigma$ level, including fluctuations and degeneracies (i.e., $\geq$84\% probability) for both MO solutions, where the precision on \Dm\ is the leading order term.
	In fact, a 5$\sigma$ measurement may be possible, at 50\% probability, within 3 years of JUNO data taking start once combined. 
	In the end, JUNO data may prescind entirely from the \Bs's \ACs\ information (minor impact), thus enabling a fully resolved pure vacuum oscillation MO measurement.
	}
	\label{Fig7}
	\vglue -0.2cm
\end{figure*}

\noindent{\bf 4.~Hypothetical MO Resolution Timeline:}
one of the main observations upon this study is that the MO could be fully resolved, maybe even comfortably, by the JUNO, \N\ and T2K combination.
The \NMOs\ solution discovery potential, considering today's favoured \dCP,  has a probability of $\geq$50\% ($\geq$84\%) for a \Dm\ precision of  up to 1.0\% (0.75\%).
In the harder \IMOs, the sensitivity may reach a mean of $\sim$5$\sigma$ potential only if the \Dm\ uncertainty was as good as $\sim$0.75\%.
Within a similar time scale, the atmospheric data is expected to add up to enable a full 5$\sigma$ resolution for both solutions.
If correct, this is likely to become the first fully resolved MO measurement and it is expected to be tightly linked to the JUNO data timeline, as described in Figure~\ref{Fig7}, which sets the timeline to be between 2026-2028.

\begin{figure*}[t]
	\vspace{0.2cm}
	\centering
	\vglue -0.3cm
	\hglue -0.2cm
        	\includegraphics[scale=0.26]{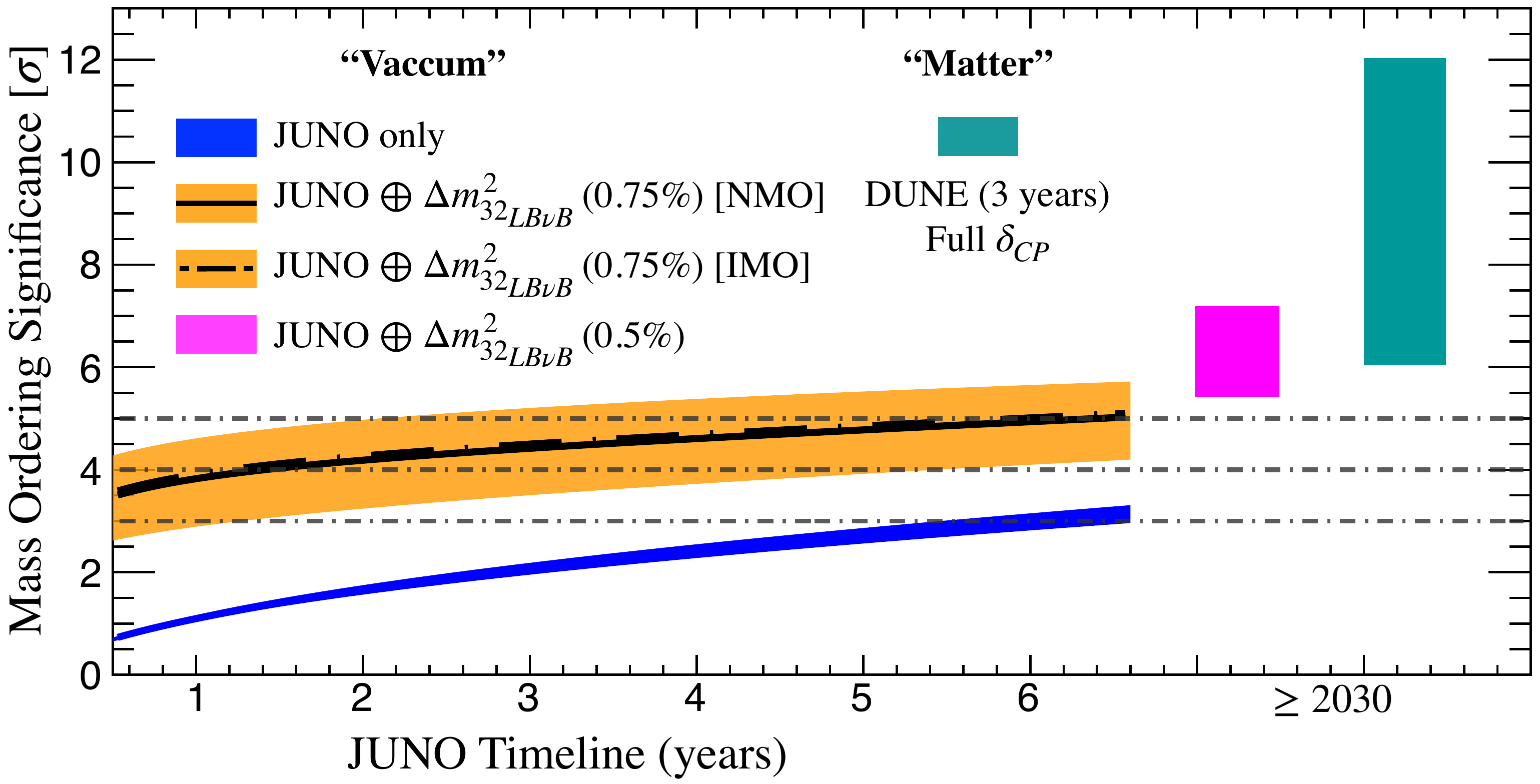}
	\vglue -0.3cm
	\caption
	{ \small 
	{\bf Different Measurements of Mass Ordering.}
	As illustrated in Figure~\ref{Fig7}, the \BIIs\ generation may provide major boosting of the JUNO sensitivity upon the boosting caused by \Dm\ precision, which is expected to be at best $\geq$0.75\%.
	The boosting effect is again illustrated as the difference between the JUNO alone (blue) and JUNO boosted (orange) curves.
	However, once the \BIIIs\ generation accelerator experiments start, we expect the \Dm\ precision to be further enhanced up to $\sim$0.5\% by both DUNE and HK.
	At this moment, the JUNO data may only exploit this \Dm\ precision to ensure a fully resolved vacuum only MO measurement (magenta), which can be compared to DUNE stand-alone measurement (green).
	Given the possible uncertainties due to experiment schedules, etc, all we can say is possible \textgreater2030.
	However, this opens for an unprecedented scenario where two as different as possible high precision MO measurements will be available to ensure the possible overall coherent of the neutrino standard phenomenology.
	Should discrepancies be seen, this may a smoking-gun evidence of the manifestation of new neutrino phenomenology.
	}
	\label{Fig8}
	\vglue -0.5cm
	\vspace{0.5cm}
\end{figure*}

Such a combined MO measurement can be regarded as a ``hybrid'' between vacuum (JUNO) and matter driven (mainly \N) oscillations.
In this context, JUNO and \N\ are, unsurprisingly, the leading experiments.
Despite holding little intrinsic MO sensitivity, T2K plays a key role by simultaneously  
a) boosting JUNO via its precise measurement of \Dm\ (similar to \N)
and 
b) aiding \N\ by reducing the possible \dCP\ ambiguity phase-space.
The \AC\ channel synergy between T2K and \N\ is expected to have very little impact.

This combined measurement relies on an impeccable 3$\nu$ data model consistency across all experiments.
Possible inconsistencies may diminish the combined sensitivity.
Since our estimate has accounted for fluctuations (typically, up to $\sim$84\% probability), those inconsistencies should amount to $\geq$2$\sigma$ effects for them to matter.
Those inconsistencies may, however, be the first manifestation of new physics \cite{Denton:2020uda}~\cite{Capozzi:2019iqn}.
Hence, this inter-experiment combination has another relevant role: to exploit the ideal MO binary parameter space solution to test for inconsistencies that may point to discoveries beyond today's standard picture.
The additional atmospherics data mentioned above, are expected to reinforce both the significance boost and the model consistency scrutiny just highlighted.

\noindent{\bf 5.~Readiness for \BIIIs:}
in the absence of any robust model-independent for MO prediction by theory and given its unique binary MO outcome, the articulation of at least two well resolved measurements appears critical for the sake of the experimental redundancy and consistency test across the field.
In the light of DUNE's unrivalled MO resolution power, the articulation of another robust MO measurement may be considered as a priority to make the most of DUNE's insight.

\noindent{\bf 6.~Vacuum versus Matter Measurements:}
since  matter effects drive all experiments but JUNO, articulating a competitive and fully resolved measurement via only vacuum oscillations has been an unsolved challenge to date.
Indeed, boosting JUNO sensitivity alone, as described in Figures~\ref{Fig4} and~\ref{Fig5}, up to $\geq$5$\sigma$ remains likely impractical in the context of \BIIs, modulo fluctuations.
However, this possibility is a priori feasible in combination with the \BIIIs\ improved precision, as shown in Figure~\ref{Fig7} and more detailed Figure~\ref{Fig8}.
The significant potential improvement in the \Dm\ precision, up to order 0.5\%~\cite{Abi:2020evt,Abe:2018uyc} may prove crucial.
Furthermore, the comparison between two fully resolved MO measurements, one using only \emph{matter effects} and one exploiting pure \emph{vacuum oscillations}, is foreseen to be one of the most insightful MO coherence tests.
So, the ultimate MO measurements comparison may be the DUNE's \ACs\ alone (even after a few years of data taking) versus a full statistics JUNO boosted by the \DCs\ of HK and DUNE improving the \Dm\ precision.
This comparison is expected to maximise the depth of the MO-based scrutiny by their stark differences in terms of mechanisms, implying dependencies, correlations, etc.
The potential for a breakthrough or even discovery, exists, should a significant discrepancy manifest here.
The expected improvement in the knowledge of \dCP\ by \BIIIs\ experiments will also play a role in facilitating this opportunity.

This observation implies that the JUNO based MO capability, despite its a priori humble intrinsic sensitivity, has the potential to play a critical role throughout the history of MO explorations.
Indeed, the first MO fully resolved measurement is likely to depend much on the JUNO sensitivity (direct and indirectly); hence JUNO should maximise (\DCS~$\geq$~9) or maintain its yield.
However, JUNO's ultimate role aforementioned may remain relatively unaffected even by a small loss in performance, providing the overall sensitivity remains sizeable (e.g. \DCS\ $\geq$ 7), as illustrated in Figures~\ref{Fig5} and \ref{Fig6}.
This is because JUNO sensitivity could still be boosted by the \Bs\ experiments by their precision on \Dm, thus sealing its legacy.
There is no reason for JUNO not to perform as planned, specially given the remarkable effort for solutions and novel techniques developed, such as the dual-calorimetry, for the control and accuracy of the spectral shape~\cite{JUNO:2020xtj}.

\noindent{\bf 7.~\Bs\ Running Strategy:}
since both \ACs\ and \DCs\ channels drive the sensitivity of \Bs\ experiments, the maximal yield for a combined MO sensitivity implies a dedicated optimisation exercise, including the role of the \dCP\ sensitivity.
Indeed, as shown, the precision on \Dm, measured via the \DCs\ channel, plays a leading role in the intrinsic MO resolution, which may even outplay the role of the \ACs\ data.
So, forthcoming beam-mode running optimisation by the \Bs\ collaborations could, and likely should, consider the impact to MO sensitivity.
In this way, if \Dm\ precision was to be optimised, this will benefit from more \emph{neutrino mode} running, leading typically to both larger signal rate and better signal-to-background ratio.
This is particularly important for T2K and HK due to their shorter baselines.
For such considerations, Figure~\ref{Fig5} might offer some guidance.

\section*{Conclusions}

This work presents a simplified calculation tuned to the latest world neutrino data, via NuFit5.0, to study the most important minimal level inter-experiment combinations to yield the earliest possible full MO resolution (i.e. $\geq$5$\sigma$).
Our first finding is that the combined sensitivity of JUNO, \N\ and T2K has the potential to yield the first resolved  measurement of MO with timeline between 2026-2028, tightly linked to the JUNO schedule since full data samples of both \N\ and T2K data are expected to be available from $\sim$2026. 
Due to the absence of any a priori MO theory based prediction and given its intrinsic binary outcome, we noted and illustrated the benefit to articulate at least two independent and well resolved ($\geq$5$\sigma$) measurements of MO.
This is even more important in the light of the decisive outcome from the next generation of long baseline neutrino beams experiments.
Such MO measurements could be exploited to over-constrain and test the standard oscillation model, thus opening for discovery potential, should unexpected discrepancies may manifest.
However, the most profound phenomenological insight using MO phenomenology is expected to be obtained by having two different and well resolved MO measurements based on only matter effects enhanced and pure vacuum oscillations experimental methodologies.
While the former is driving most of the field, the challenge was to be able to articulate the latter, so far considered as impractical.
Hence, we here describe the feasible path to promote JUNO's MO measurement to reach a robust $\geq$5$\sigma$ resolution level without compromising its unique vacuum oscillation nature by exploiting the next generation long baseline neutrino beams disappearance channel's ability to reach a precision of $\leq$0.5\% on \Dm.

\section*{Acknowledgment}
{\small
	Much of this work was originally developed in the context of our studies linked to the PhD thesis of Y.H. (APC and IJC laboratories) and to the scientific collaboration between H.N. (in sabbatical at the IJC laboratory) and A.C.
	Y.H. and A.C. are grateful to the CSC fellowship funding of the PhD fellow of Y.H. 
	H.N. acknowledges CAPES and is especially thankful to CNPq and IJC laboratory for their support to his  sabbatical.
	A.C. and L.S. acknowledge the support of the
	P2IO LabEx (ANR-10-LABX-0038) in the framework ``\emph{Investissements d'Avenir}'' (ANR-11-IDEX-0003-01 -- Project ``NuBSM'') managed by the Agence Nationale de la Recherche (ANR), France, where our developments are framed within the \textit{neutrino inter-experiment synergy} working group.
	A.C. would like to thank also St\'ephane Lavignac for useful comments and suggestions as feedback on the manuscript. The authors are grateful to JUNO's internal reviewers who ensured that the information included in this manuscript about that experiment is consistent with its official position as conveyed in its publications.
	We would like to specially thank the NuFit5.0 team 
	(Ivan~Esteban, 
	Concha~Gonzalez-Garcia, 
	Michele~Maltoni, 
	Thomas~Schwetz 
	and 
	Albert~Zhou) for their kindest aid and support to provide dedicated information from their latest NuFit5.0 version.
	We also would like Concha~Gonzalez-Garcia and Fumihiko~Suekane for providing precious feedback on a short time scale and internal review of the original manuscript.
}

\bibliographystyle{unsrt}
\bibliography{references_ResolveMO4.0.bib}

\newpage
\section*{APPENDICES}
\setcounter{figure}{0}
\renewcommand{\thefigure}{A\arabic{figure}}
\setcounter{table}{0}
\renewcommand{\thetable}{A\arabic{table}}

\subsection*{A.~Empirical Reproduction of the \CS~Function for the \BIIs\ Experiments}
\label{Appendix-T2K-\N}

In this section, we shall detail how we computed the number of events for T2K and \N.
For a constant matter density, without any approximation, appearance oscillation 
probability for given baseline $L$ and neutrino energy $E$, 
can be expressed~\cite{Kimura:2002wd} as
\begin{eqnarray}
P(\nu_\mu \to \nu_e) 
&= & 
a_\nu + b_\nu \cos \delta_\text{\tiny CP} + c_\nu \sin \delta_\text{\tiny CP},
\nonumber \\
P(\bar{\nu}_\mu \to \bar{\nu}_e) 
&= & 
a_{\bar{\nu}} + b_{\bar{\nu}}  \cos \delta_\text{\tiny CP} + c_{\bar{\nu}} \sin \delta_\text{\tiny CP}, 
\end{eqnarray}
where 
$a_\nu$, $b_\nu$, $c_\nu$, $a_{\bar{\nu}}$, $b_{\bar{\nu}}$ and $c_{\bar{\nu}}$ 
are some factors which depend on the mixing parameters
($\theta_{12}$, $\theta_{23}$,  $\theta_{13}$, $\delta m^2_{21}$ and $\Delta m^2_{32}$),
$E$, $L$ as well as the matter density. 
This implies that, even after taking into account the neutrino flux spectra,
cross sections, energy resolution, detection efficiencies, and so on, 
which depend on neutrino energy, and after performing integrations 
over the true and reconstructed
neutrino energies, 
the expected number of $\nu_e$ ($\bar{\nu}_e$) appearance events, 
$N_{\nu_e}$ $(N_{\bar{\nu}_e})$, 
for a given experimental exposure (running
time) have also the similar $\delta_\text{\tiny CP}$ dependence as, 
\begin{eqnarray}
N_{\nu_e}
&= & 
n_0  + n_c \cos \delta_\text{\tiny CP} + n_s\sin \delta_\text{\tiny CP},
\nonumber \\
N_{\bar{\nu}_e} 
&= & 
\bar{n}_0 + \bar{n}_c  \cos \delta_\text{\tiny CP} + \bar{n}_s \sin \delta_\text{\tiny CP}, 
\label{eq:accelerator_event_number}
\end{eqnarray}
where $n_0$, $n_c$,  $n_s$, $\bar{n}_0$, $\bar{n}_c$ and $\bar{n}_s$ 
are some constants which depend not only on mixing parameters but 
also on experimental setups. 
Assuming that background (BG) events do not depend
(or depend very weakly) on $\delta_\text{\tiny CP}$,  
the constant terms $n_0$ and $\bar{n}_0$ in Eq.~\eqref{eq:accelerator_event_number}
can be divided into the signal
contribution and BG one as $n_0 = n_0^\text{sig} + n_0^\text{BG}$
and $\bar{n}_0 = \bar{n}_0^\text{sig} + \bar{n}_0^\text{BG}$, as an approximation.

\begin{figure*}
 \vspace{-0.3cm}
 \begin{tabular}{cc}
 \begin{minipage}[t]{0.5\hsize}
   \hglue -1.cm
   \centering
\includegraphics[scale=0.35]{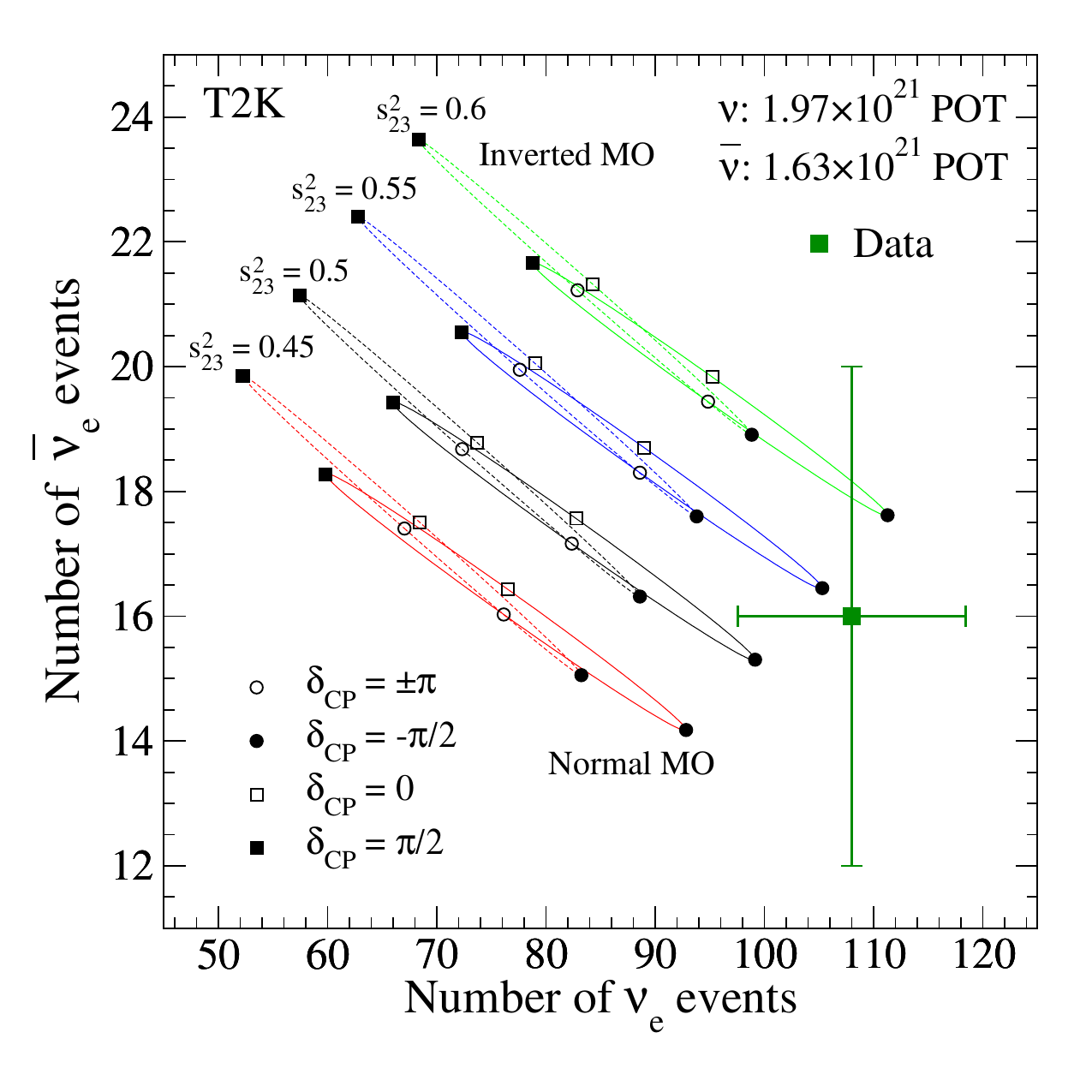}
 \end{minipage} &
\begin{minipage}[t]{0.46\hsize}
  \hglue -1.cm
  \centering
  \includegraphics[scale=0.35]{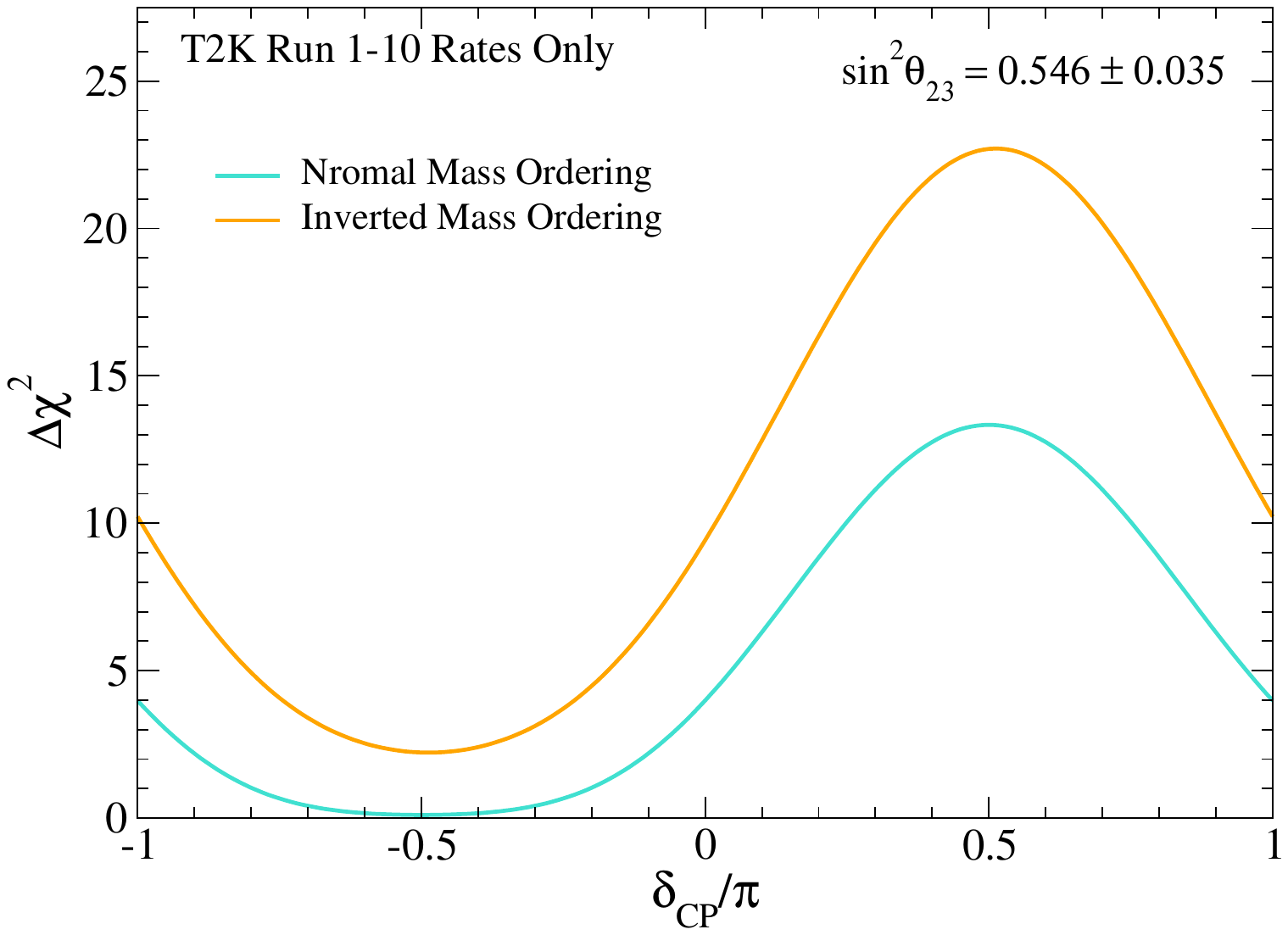}
      \end{minipage}
    \end{tabular}
    \vspace{-0.5cm}
	\caption
	{ \small {\bf Reproduction of T2K Bi-Rate and CP Sensitivity Results as of Neutrino 2020 Conference.}
	Left panel is the bi-rate plot which shows simultaneously the expected number of events for $\nu_e$ and $\bar{\nu}_e$ by varying continuously the values of $\delta_\text{\tiny CP}$ from $-\pi$ to $\pi$, indicated by the solid and dashed curves (ellipses) for 4 different values of $s^2_{23} = \sin^2 \theta_{23}$ indicated in the legend for the exposure of 2.0(1.6) $\times 10^{21}$  POT for neutrino (anti-neutrino) mode. 
	The point corresponding to the latest T2K data reported at Neutrino 2020 Conference~\cite{Ref_T2K@Nu2020} is indicated by the solid dark green square with 1$\sigma$ error bars. 
	Right panel shows the $\Delta \chi^2$ obtained by fitting the data using the $\chi^2$ function given in Eq.~\eqref{eq:chi2_LBnuB}. 
	}
\label{Fig9}
\end{figure*}

\begin{figure*}
 \vspace{-0.3cm}
 \begin{tabular}{cc}
 \begin{minipage}[t]{0.5\hsize}
\centering
\hglue -1cm
\includegraphics[scale=0.35]{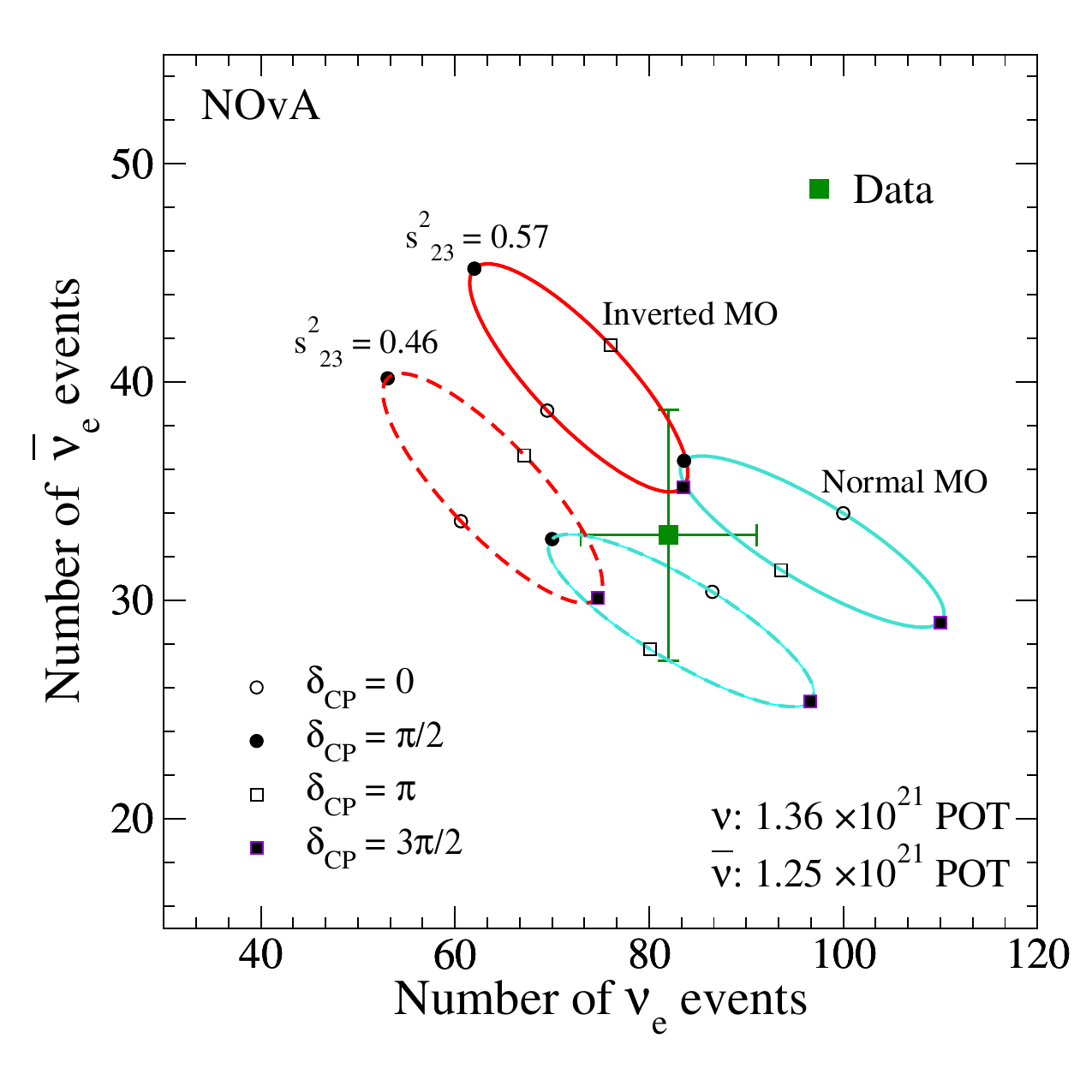}
      \end{minipage} &
      \begin{minipage}[t]{0.6\hsize}
\hglue -3.cm
\centering
        \includegraphics[scale=0.35]{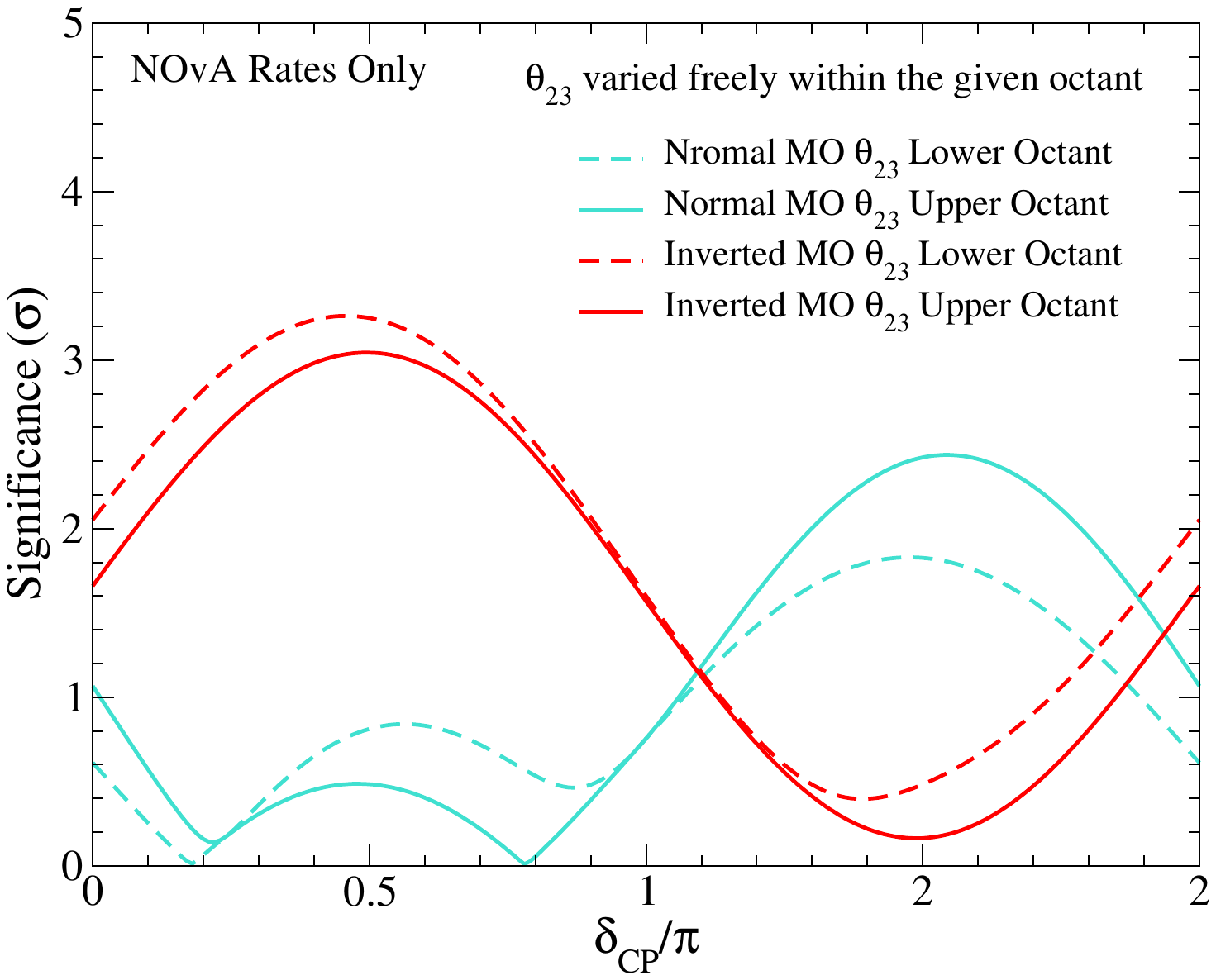}
      \end{minipage}
    \end{tabular}
 \vspace{-0.5cm}
 \caption
	{ \small {\bf Reproduction of \N\ Bi-Rate and CP Sensitivity Results as of Neutrino 2020 Conference.}
	Left panel is the bi-rate plot which shows simultaneously the expected number of events for $\nu_e$ and $\bar{\nu}_e$ by varying continuously the values of $\delta_\text{\tiny CP}$ from $-\pi$ to $\pi$, indicated by the solid and dashed curves (ellipses) for 4 different values of $s^2_{23} = \sin^2 \theta_{23}$ indicated in the legend for the exposure of 1.4 (1.3) $\times 10^{21}$  POT for neutrino (anti-neutrino) mode. 
	The point corresponding to the latest \N\,
	 data reported at Neutrino 2020 Conference~\cite{Ref_NOVA@Nu2020} is indicated by the solid dark green square with 1$\sigma$ error bars. 
	Right panel shows the significance $\sqrt{\Delta \chi^2}$ obtained by fitting the data using the $\chi^2$ function given in Eq.~\eqref{eq:chi2_LBnuB} but by setting $\chi^2_\text{pull}(\sin^2 \theta_{23})$ =0.
	}
	\label{Fig10}
\end{figure*}

In Table~\ref{table:t2k_nova_coefficients}, we provide the numerical values of these coefficients which can reproduce quite well the expected number of events shown in the plane spanned by ${N^\text{obs}_{\nu_e}}$ and $N^\text{obs}_{\bar{\nu}_e}$, often called bi-rate plots, 
found in the presentations by 
T2K~\cite{Ref_T2K@Nu2020} and \N~\cite{Ref_NOVA@Nu2020}
at Neutrino 2020 Conference, for their corresponding
accumulated data (or exposures). 
We show in the left panels of Figures~\ref{Fig9} and \ref{Fig10}, respectively, for T2K and \N, the bi-rate plots which were reproduced by using the values given in Table~\ref{table:t2k_nova_coefficients}.
Our results are in excellent agreement with the ones shown by the collaborations~\cite{Ref_T2K@Nu2020,Ref_NOVA@Nu2020}.

The $\chi^2$ function for the appearance channel (AC), 
for a given LB$\nu$B experiment, T2K or \N, 
which is based on the total number of events, 
is simply defined as follows, for each MO, 
\begin{eqnarray}
{\chi^2}_{\text{\tiny LB$\nu$B}}^{\text{ AC}}
&&
\hskip -0.7cm
\equiv 
\min_{s^2_{23},\,\delta_\text{\tiny CP}}
\left[
\frac{ 
( N^\text{obs}_{\nu_e}
- N^\text{theo}_{\nu_e}(s^2_{23},\delta_\text{\tiny CP}) )^2
}
{N^\text{obs}_{\nu_e}}
\right. \nonumber \\
&& \hskip -2.0cm +
\left.\frac{(N^\text{obs}_{\bar{\nu}_e} -
N^\text{theo}_{\bar{\nu}_e}(s^2_{23},\delta_\text{\tiny CP}) )^2}
{N^\text{obs}_{\bar{\nu}_e}}
+ \chi^2_\text{pull}(\sin^2 \theta_{23})\right],
\label{eq:chi2_LBnuB}
\end{eqnarray}
where $N^\text{obs}_{\nu_e}$ ($N^\text{obs}_{\bar{\nu}_e}$) 
is the number of observed (or to be observed) $\nu_e$ ($\bar{\nu}_e$)
events, 
and $N^\text{theo}_{\nu_e}$ ($N^\text{theo}_{\bar{\nu}_e}$) 
are the corresponding theoretically expected numbers (or prediction), 
and 
\begin{eqnarray}
\chi^2_\text{pull}(\sin^2 \theta_{23}) \equiv
\left(\frac{\sin^2 \theta_{23}^\text{\ true} -\sin^2 \theta_{23} }{\sigma(\sin^2 \theta_{23})}\right)^2.
\label{eq:chi2_th23_pull}
\end{eqnarray}
Note that the number of events in Eq.~\eqref{eq:chi2_LBnuB}
include also background events.

\begin{table}[H]
\begin{center}%
\begin{tabular}{c|c|c|c|c} 
\hline \hline
& ${n}_0^\text{sig}$/$\bar{n}_0^\text{sig}$
& $n_0^\text{BG}$/$\bar{n}_0^\text{BG}$  
& $n_c$/$\bar{n}_c$ & $n_s$/$\bar{n}_s$ \\
\hline 
T2K $\nu$ NMO  & 68.6 & 20.2 & 0.2 & -16.5\\
T2K $\bar{\nu}$ NMO & 6.0& 12.5 & 0.2 & 2.05\\
\hline 
T2K $\nu$ IMO  & 58.1 & 20.2 & 0.7 & -15.5\\
T2K $\bar{\nu}$ IMO & 14.0 & 6.0 & 0.05 & 2.40\\
\hline 
\N $\nu$ NMO & 70.0 & 26.8 & 3.2 & -13.2 \\
\N $\bar{\nu}$ NMO & 18.7  & 14.0 & 1.3 & 3.7 \\
\hline 
\N $\nu$ IMO& 45.95 & 26.8 & -3.25 & -10.75 \\
\N $\bar{\nu}$ IMO & 26.2  & 14.0 & -1.5 & 5.0 \\
\hline \hline
\end{tabular}%
\end{center}
\caption{\small
	{\bf \N\ and T2K Oscillation Probability Empirical Parametrisation as of Neutrino 2020 Conference.}
	The numerical values of the factors appear in Eq.~(\ref{eq:accelerator_event_number}) are shown, which were adjusted to approximately agree with what have been presented by T2K~\cite{Ref_T2K@Nu2020} and \N~\cite{Ref_NOVA@Nu2020}.  
	These numbers correspond to the exposures of 2.0(1.6) $\times 10^{21}$ protons on target (POT) for $\nu$ ($\bar{\nu}$) mode of T2K and 1.4(1.3) $\times 10^{21}$ POT for $\nu$ ($\bar{\nu}$) mode of \N~ experiments. 
	The 3 factors ${n}_0^\text{sig}$, $n_c$ and $n_s$ correspond to the case where $\sin^2 \theta_{23} = 0.55$ (0.57) for T2K (\N) and they scale as ${n}_0^\text{sig} \propto \sin^2 \theta_{23}$ and ${n}_c, n_s \propto \sin^2 2\theta_{23}$ as $\theta_{23}$ varies. 
	The values of $\Delta m^2_{32}$ are fixed to $\Delta m^2_{32} = 2.49 (-2.46) \times 10^{-3}$ eV$^2$ for NMO (IMO) for T2K~\cite{Ref_T2K@Nu2020} and $\Delta m^2_{32} = 2.40(-2.44)\times 10^{-3}$ eV$^2$ for NMO for \N~\cite{Ref_NOVA@Nu2020,Ref_NOVA@ICHEP2020}. 
	}
\label{table:t2k_nova_coefficients}
\end{table}

Using the number of events given in Eq.~\eqref{eq:accelerator_event_number} with values of coefficients given in Table \ref{table:t2k_nova_coefficients} we performed a fit to the data recently reported by T2K at Neutrino 2020 Conference~\cite{Ref_T2K@Nu2020} just varying $\sin^2 \theta_{23}$ and $\delta_\text{\tiny CP}$ and could reproduce rather well the $\Delta \chi^2$ presented by T2K in the same conference mentioned above, as shown in the right panel of Figure~\ref{Fig9}.
We have repeated the similar exercises also for \N~
 and obtained the results, shown in the right panel of Figure~\ref{Fig10},
 which are reasonably in agreement with 
what was presented by \N~at 
at Neutrino 2020 Conference~\cite{Ref_NOVA@Nu2020}.
In the case of \N~ the agreement is slightly worse as compared to the case of T2K. 
  We believe that this is because, for the results shown in Figure~\ref{Fig10}, unlike the case of T2K, we did not take into account the $\theta_{23}$ constraint by \N\ (or we have set $\chi_\text{pull}$ in Eq.~\eqref{eq:chi2_th23_pull} equals to zero) as this information was not reported in~\cite{Ref_NOVA@Nu2020}.

We note that for this part of our analysis, we considered only the dependence of $\sin^2 \theta_{23}$ and $\delta_\text{\tiny CP}$ and ignore the uncertainties of all the other
mixing parameters as we are computing the number of events in an approximated way, as described above, by taking into account only the variation due to $\sin^2 \theta_{23}$ and $\delta_\text{\tiny CP}$ with all the other parameters fixed (separately by T2K~\cite{Ref_T2K@Nu2020} and \N~\cite{Ref_NOVA@Nu2020} collaborations) to some values which are  close to the values given in Table~\ref{table:mixing-parameters-nufit5.0}.

In particular, we neglected the uncertainty of $\Delta m^2_{32}$ in the LB$\nu$B AC part analysis when it is combined with JUNO plus LB$\nu$B DC part analysis to obtain our final boosted MO sensitivities. 
Strictly speaking, $\Delta m^2_{32}$ must be varied simultaneously (in a synchronised way) in the $\chi^2$ defined in Eq.~\eqref{eq:chi2_LBnuB_AC} when it is combined with  the $\chi^2$ defined in Eq.~\eqref{eq:chi2_juno_plus_pull}. 
However, in our analysis, we simply add $\Delta \chi^2$ obtained from our simplified LB$\nu$B AC simulation which ignored $\Delta m^2_{32}$ uncertainty, to the JUNO's boosted $\chi^2$ (described in detail in the Appendix~C).  
This can be justified by considering that a variation of $\Delta m^2_{32}$ of about $\sim$ 1\% imply only a similar magnitude of variations in the appearance oscillation probabilities, which would be significantly smaller than the statistical uncertainties of LB$\nu$B-II AC mode, which are expected to reach at most the level or $\sim$5\% or larger even in our future projections for T2K and \N.

For the MO resolution sensitivity shown in Figure~\ref{Fig2} and used for our analysis throughout this work, we define the $\Delta {\chi^2}$ (labeled as  
$\Delta {\chi^2}_\text{LB$\nu$B}^\text{ AC}$), as 
\begin{eqnarray}
\Delta {\chi^2}_\text{LB$\nu$B}^\text{ AC} (\text{MO})
&&\equiv \nonumber \\ 
&&
\hskip -3.8cm \pm
\min_{\sin^2 \theta_{23},\,\delta_\text{\tiny CP}}
\left[
{\chi^2}_{\text{\tiny LB$\nu$B}}^{\text{ AC}}(\text{IMO})
-{\chi^2}_{\text{\tiny LB$\nu$B}}^{\text{ AC}}(\text{NMO})
\right],
\label{eq:chi2_LBnuB_AC}
\end{eqnarray}
where +(-) sign corresponds to the case where 
the true MO is normal (inverted),
and 
${\chi^2}_{\text{\tiny LB$\nu$B}}^{\text{ AC}}$
is computed as defined in Eq.~\eqref{eq:chi2_LBnuB}
but with $N^\text{obs}_{\nu_e (\bar{\nu}_e)}$
replaced by the theoretically expected ones 
for given values of assumed true values 
of $\theta_{23}$ and $\delta_\text{\tiny CP}$. 
In practice, since we do not consider the effect of fluctuation for this part of our analysis, ${{\chi^2}_{\text{\tiny LB$\nu$B}}^{\text{ AC}}}_\text{ min}= 0$ by construction for true MO.
We note that when T2K and \N\ are combined, 
some enhancement of sensitivities in the 
positive (negative) $\delta_\text{\tiny CP}$ region for NMO (IMO)
occur (see light green curves in Figure \ref{Fig6}). 
This is because that in these $\delta_\text{\tiny CP}$ ranges,
T2K and \N\, data can not
be simultaneously fitted very well by 
using the common $\delta_\text{\tiny CP}$ for the wrong MO, 
leading to an increase of $\Delta \chi^2$.

For simplicity, for our future projection, we simply increase by a factor of 3 both T2K and \N\ exposures, to the coefficients given in Table~\ref{table:t2k_nova_coefficients} for both $\nu$ and $\bar{\nu}$ channels.
This corresponds approximately to 8.0 (6.4)$\times 10^{21}$ POT for T2K $\nu\,(\bar{\nu})$ mode and to 4.1 (3.8)$\times 10^{21}$ POT for \N~$\nu\,(\bar{\nu})$ mode, to reflect roughly the currently considered final exposures for T2K~\cite{Ichikawa:2020} 
($\simeq 10\times 10^{21}$ POT in total for $\nu$ and $\bar{\nu}$)
and \N~\cite{Ref_NOVA@ICHEP2020}
($\simeq 3.2\times 10^{21}$ POT each for $\nu$ and $\bar{\nu}$).
This approach implies that our calculation does not consider future unknown optimisations on the $\nu\,(\bar{\nu})$ mode running.

\subsection*{B.~\Bs~Disappearance MO Sensitivity}

\begin{figure*}[t!]
    \vspace{-1.cm}
    \hspace{0.3cm}
	\centering
	\hglue -0.4cm
	\includegraphics[scale=0.55]{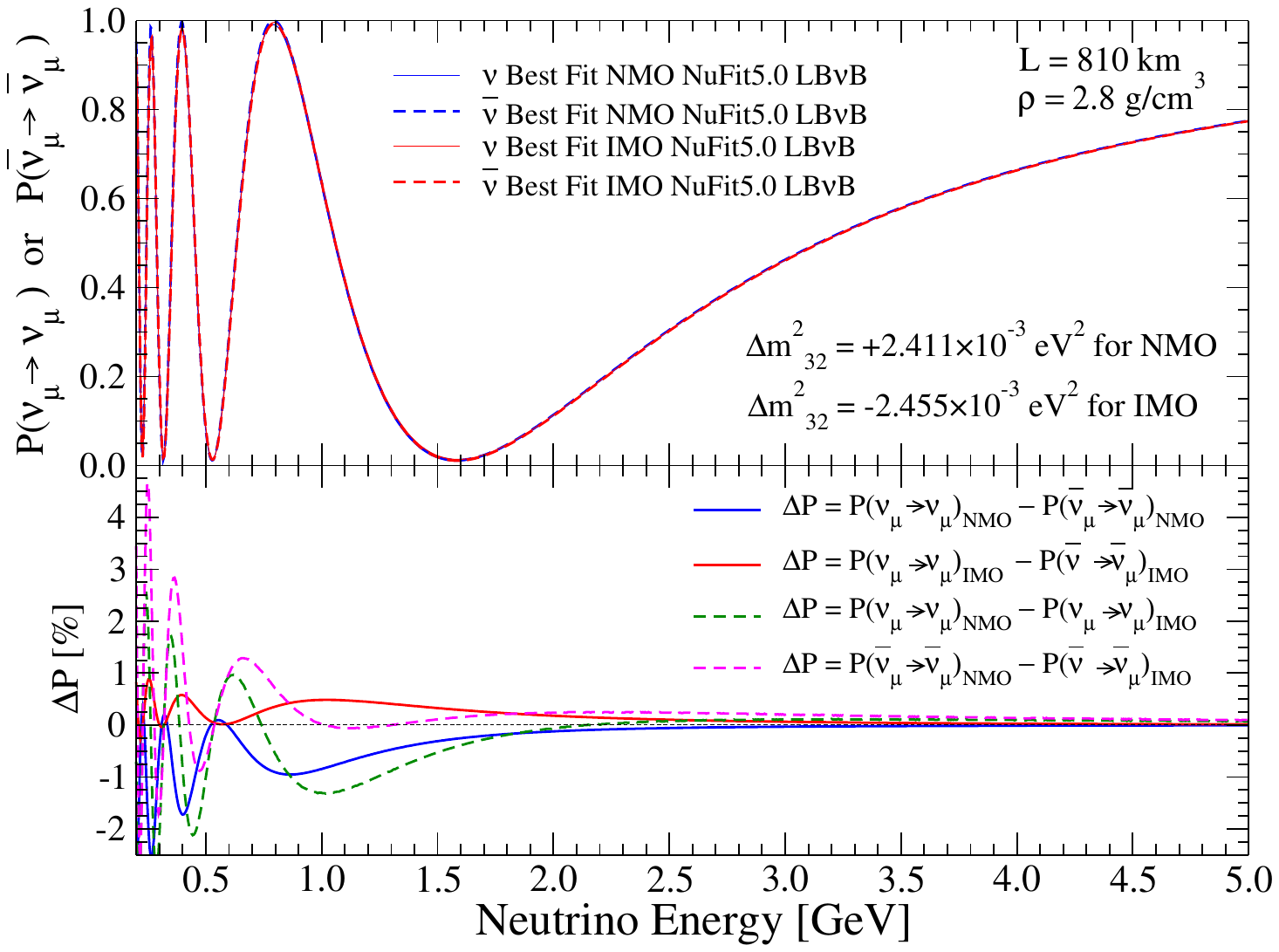}
	\vspace{-1.cm}
	\caption
	{ \small {\bf \Bs\ Survival Probability Mass Ordering Dependence.}
	In the upper panel, the $\nu_\mu \to \nu_\mu$ and $\bar{\nu}_\mu \to \bar{\nu}_\mu$ survival probabilities computed by using the mixing parameters found in Table~\ref{table:mixing-parameters-nufit5.0} are shown for NMO (solid and dashed red curves) and IMO (solid and dashed blue curves) as a function of neutrino energy, as indicated in the legend. 
	In the lower panel, the differences of these probabilities are shown in percent. 
	}
	\label{Fig11}
\end{figure*}

In the upper panel of Figure~\ref{Fig11}, we show the 
4 curves of survival oscillation probabilities, 
$P(\nu_\mu \to \nu_\mu)$ and $P(\bar{\nu}_\mu \to \bar{\nu}_\mu$)
for NMO and IMO, which were obtained by using the best 
fitted parameters in NuFit5.0 given
in Table~\ref{table:mixing-parameters-nufit5.0}
for the baseline corresponds to \N ($L=810$ km) and with
the matter density of $\rho = 2.8$ g/cm$^3$. 
The NMO and IMO cases are shown, respectively, by blue
and red colours whereas the cases for $\nu$ and $\bar{\nu}$
are shown, respectively, by solid thin and dashed thick curves. 
We observe that all of these 4 curves coincide very well with each other, so differences are very small.
In the lower panel of the same Figure~\ref{Fig11}, we show the differences of these curves, between $\nu$ and $\bar{\nu}$ channels for both NMO and IMO, as well as between NMO and IMO for both $\nu$ and $\bar{\nu}$, as indicated in the legend. 
We observe that the differences of these oscillation probabilities are $\leq$1\% for the energy range relevant for \N. 

Two points can be highlighted. 
First, the fact that the differences between neutrino and anti-neutrino are quite small implies that the matter effects are very small in these channels, hence determining MO by using matter effects based only on LB$\nu$B~DC would be almost impossible.
And second, the fact that the curves for NMO and IMO agree very well implies that the absolute values of the effective mass squared differences, called $\Delta m^2_{\mu\mu}$, defined in Eq.~\eqref{eq:dm_mumu} in Appendix~D, which correspond to NMO and IMO cases, should be similar. 
Indeed, by using the values given in Table~\ref{table:mixing-parameters-nufit5.0}, we obtain $\Delta m^2_{\mu\mu} = 2.422 (-2.431)\times 10^{-3}$eV$^2$ for NMO (IMO) exhibiting a small $\sim 0.4$\% difference. 
In other words, for each channel, $\nu$ and $\bar{\nu}$, there are two {\it degenerate} solutions, one corresponds to NMO and the other, to IMO, which give in practice the same survival probabilities.
We stress that this degeneracy can not be resolved by considering LB$\nu$B experiment with DC alone. 

\subsection*{C.~Analytic Understanding of Synergy between JUNO and
  LB$\nu$B based experiments}

In this section, we shall detail the relation between true and false $\Delta m^2_{32}$ solutions in the case of JUNO and \Bs, as they are different.
This difference is indeed exploited as the main numerical quantification behind the \DCSB\ term which was schematically illustrated in Figure~\ref{Fig3} in the main text
and will be further quantified in Figure~\ref{FigA4} to be shown in this appendix.

\subsubsection*{C.1~JUNO Relation between True-False $\Delta m^2_{32}$}

The $\bar{\nu}_e \to \bar{\nu}_e$ survival probability in vacuum 
can be expressed as~\cite{Minakata:2007tn}
\begin{eqnarray}
\label{eq:prob_ee}
\hskip -0.5cm P_{\bar{\nu}_{e}\rightarrow\bar{\nu}_{e}} 
&=& 1-c^4_{13}\sin^{2}2\theta_{12}\sin^{2}\Delta_{21} 
-\frac{1}{2}\sin^{2}2\theta_{13}  \nonumber \\
& & \hskip -2cm 
\times \left[1- \sqrt{1-\sin^{2}2\theta_{12}\sin^{2}\Delta_{21}}
\cos(2|\Delta_{ee}|\pm\phi)\right],
\end{eqnarray}
where the notation $c_{ij} \equiv \cos \theta_{ij}$
and $s_{ij} \equiv \sin \theta_{ij}$ is used, and 
$\Delta_{ij}\equiv\Delta m_{ij}^{2}L/4E$, $L$ and $E$ are,
respectively, the baseline and the neutrino energy,
and the effective mass squared difference 
$\Delta m^2_{ee}$ is given by~\cite{Nunokawa:2005nx} 
\begin{eqnarray}
\Delta m^2_{ee} \equiv 
c^2_{12}\Delta m^2_{31} + s^2_{12}\Delta m^2_{32}
= \Delta m^2_{32} + c^2_{12}\Delta m^2_{21}, 
\label{eq:dmee}
\end{eqnarray}
and $\phi$ is given by
\begin{eqnarray}
\tan\phi=\frac
{c_{12}^{2}\sin(2s_{12}^{2}\Delta_{21})-s_{12}^{2}\sin(2c_{12}^{2}\Delta_{21})}
{c_{12}^{2}\sin(2s_{12}^{2}\Delta_{21})+s_{12}^{2}\sin(2c_{12}^{2}\Delta_{21})},
\end{eqnarray}
where $\phi \simeq 0.36$ radian $\simeq 0.11 \pi$ for
$s^2_{12} =0.304$ and $\delta m^2_{21} = 7.42\times 10^{-5}$ eV$^2$.
The +(-) sign in front of $\phi$ in Eq.~\eqref{eq:prob_ee} 
corresponds to the normal (inverted) mass ordering. 

Upon data analysis, JUNO will obtain two somewhat different values of $\Delta m^2_{32}$ corresponding to NMO and IMO, which we call 
$\Delta {m^2_{32}}_\text{\tiny\ JUNO}^\text{\tiny NMO}$ 
and 
$\Delta {m^2_{32}}_\text{\tiny\ JUNO}^\text{\tiny IMO}$
where one of them should correspond (or closer) to the true solution. 
It is expected that by considering 
$\Delta_{ee}^\text{\tiny NMO} + \phi =\Delta_{ee}^\text{\tiny IMO}-\phi$,
they are approximately related by 
\begin{eqnarray}
\Delta {m^2_{32}}_\text{\tiny\ JUNO}^\text{\tiny IMO}
\simeq - \Delta {m^2_{32}}_\text{\tiny\ JUNO}^\text{\tiny NMO} -
2c^2_{12}\delta m^2_{21} - \delta m^2_\phi,
\label{eq:dm2_32_relation_juno} 
\end{eqnarray}
where the approximated value of $\delta m^2_\phi$ can be 
estimated by choosing the average representative energy of reactor neutrinos ($\sim $4 MeV) as 
\begin{eqnarray}
 \delta m^2_\phi \equiv \frac{4E}{L}\phi
\simeq 2.1 \times 10^{-5} \left(\frac{E}{4\ \text{MeV}}\right) 
\label{eq:dm2_phi} 
\text{eV}^2.
\end{eqnarray}
We found that for a given assumed true value of
$\Delta {m^2_{32}} = 2.411 \times 10^{-3}$ eV$^2$ (corresponding to NMO), 
we can reproduce very well the false value of 
$\Delta {m^2_{32}} = -2.53 \times 10^{-3}$ eV$^2$ (corresponding to
IMO) obtained by a $\chi^2$ fit if we use $E=4.4$ MeV 
in Eqs.~\eqref{eq:dm2_32_relation_juno} and \eqref{eq:dm2_phi}. 
The relation between true and false 
$\Delta {m^2_{32}}$ for JUNO is illustrated by the vertical 
black dashed and black solid lines 
in Figure~\ref{FigA4} (b) and (d).

\subsubsection*{C.2~\Bs\ Relation between True-False $\Delta m^2_{32}$} 
For \Bs\ experiments like T2K and \N\, the $L/E$ are such that 
$|\Delta_{31}| \sim |\Delta_{32}|  \sim \pi/2$. 
From the disappearance channels
$\nu_\mu \to \nu_\mu$ 
and 
$\bar{\nu}_\mu \to \bar{\nu}_\mu$,
it is possible to measure precisely the effective mass squared difference $\Delta m^2_{\mu\mu}$ whose value is independent of the MO.
In terms of fundamental mixing and oscillation parameters, $\Delta m^2_{\mu\mu}$ can be expressed, with very good approximation, as  
~\cite{Nunokawa:2005nx}, 
\begin{eqnarray}
&&  \hskip -1cm \Delta m^2_{\mu\mu}\equiv \Delta m^2_{32}   +
  \nonumber \\
&&(s^2_{12} + \cos\delta_\text{\tiny CP} s_{13}\sin2\theta_{12}\tan \theta_{23})
\delta m^2_{21}.
\label{eq:dm_mumu} 
\end{eqnarray}
From this relation, one can extract two possible values of $\Delta m^2_{32}$
corresponding to two different MO as
\begin{eqnarray}
\Delta{m^2_{32}}^\text{\tiny MO}_\text{ LB$\nu$B} 
&=& 
+(-) |\Delta m^2_{\mu\mu}| - \nonumber \\
&&
\hskip -2cm (s^2_{12} + \cos\delta_\text{\tiny CP}^\text{\tiny MO} 
s_{13}^\text{\tiny MO} \sin2\theta_{12}
\tan\theta_{23}^\text{\tiny MO} )
\delta m^2_{21}, 
\label{eq:dm2_32_dm2_mumu_relation}
\end{eqnarray}
where superscript MO implies either NMO or IMO, and + and - sign 
correspond, respectively, to NMO and IMO. 
Note that the best fitted values for the mixing and oscillation parameters, with the exception of solar parameters
$\theta_{12}$ and $\delta m^2_{21}$, can be different in the NMO and IMO scenarios.
Eq.~\eqref{eq:dm2_32_dm2_mumu_relation} can be rewritten as
\begin{eqnarray}
\Delta{m^2_{32}}^\text{\tiny IMO}_\text{\tiny LB$\nu$B} 
=-\Delta{m^2_{32}}^\text{\tiny NMO}_\text{\tiny LB$\nu$B} 
 -\delta m^2_{21} \left\{2s^2_{12} 
+ \sin2\theta_{12} \right.\nonumber \\
&&
\hskip -9cm 
\left.
(
\cos\delta_\text{\tiny CP}^\text{\tiny NMO} s_{13}^\text{\tiny NMO} 
\tan\theta_{23}^\text{\tiny NMO} 
+ \cos \delta_\text{\tiny CP}^\text{\tiny IMO} s_{13}^\text{\tiny IMO}
\tan\theta_{23}^\text{\tiny IMO} ) \right\}
\nonumber \\
\simeq -\Delta{m^2_{32}}^\text{\tiny NMO}_\text{\tiny LB$\nu$B} 
 -\delta m^2_{21} \left\{ 2s^2_{12} 
+ \sin2\theta_{12} \right.\nonumber \\
&&
\hskip -6.6cm 
\left.\times s_{13} \tan\theta_{23}(
\cos\delta_\text{\tiny CP}^\text{\tiny NMO} 
+ \cos \delta_\text{\tiny CP}^\text{\tiny IMO} 
)
\right\},
\label{eq:dm2_32_relation_acc}
\end{eqnarray}
where in the last line of the above equation, some simplifications 
were considered based on the fact that best fitted values of 
$\sin^2\theta_{13}$ 
and 
$\sin^2\theta_{23}$ 
in recent global analysis \cite{NuFit5.0} are similar for both MO solutions. 
By using the relation given in Eq.~\eqref{eq:dm2_32_relation_acc}, for a given assumed true value of $\Delta{m^2_{32}}$ (common for all experiments) we obtain the yellow colour
bands shown in Figure~\ref{FigA4} (b) and (d).

\subsubsection*{C.3~Boosting Synergy Estimation}

\begin{figure*}
  \centering
	\hglue -0.1cm
	\includegraphics[scale=0.44]{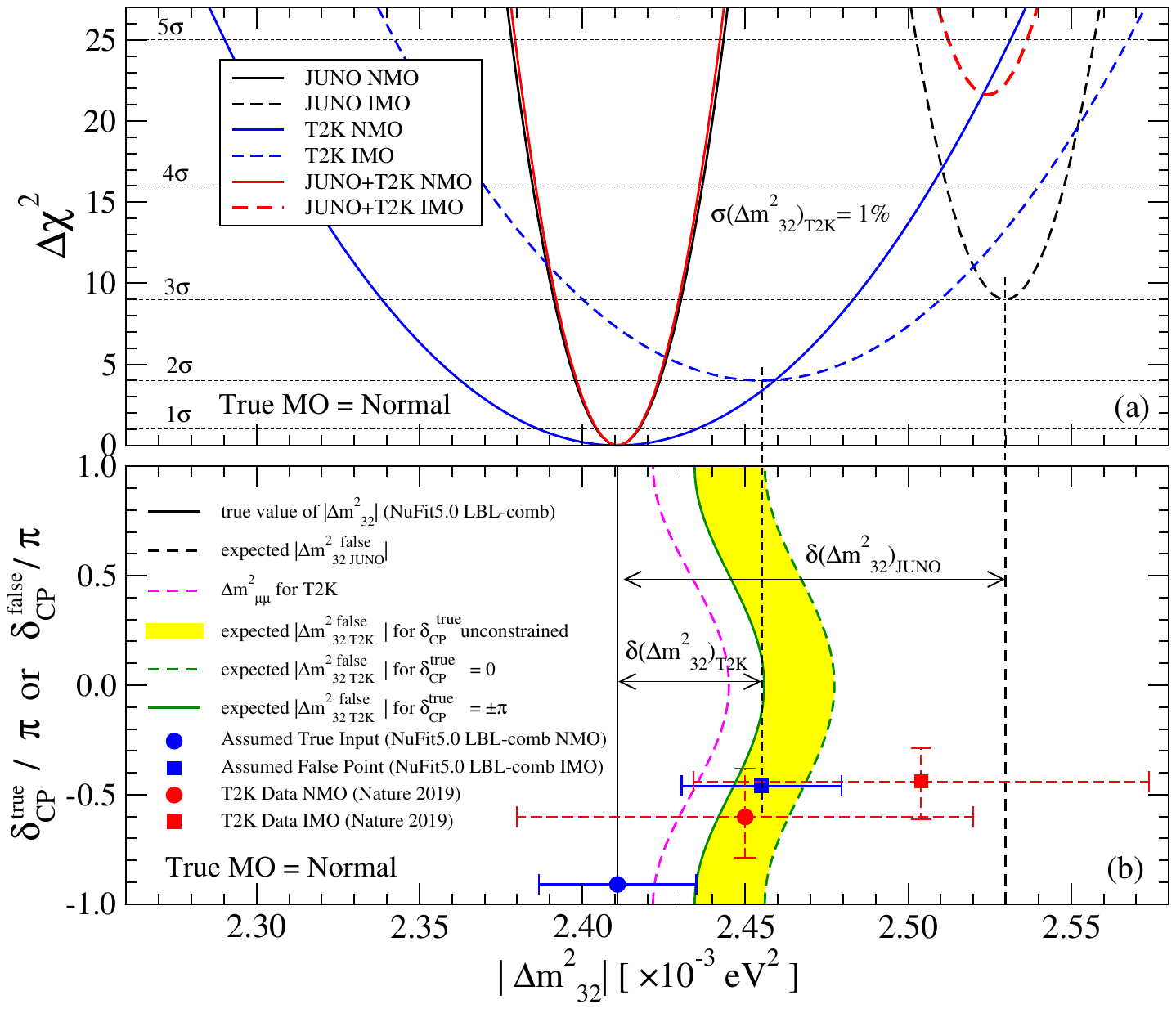}
	\vskip -0.3cm
	\hglue -0.1cm
	\includegraphics[scale=0.44]{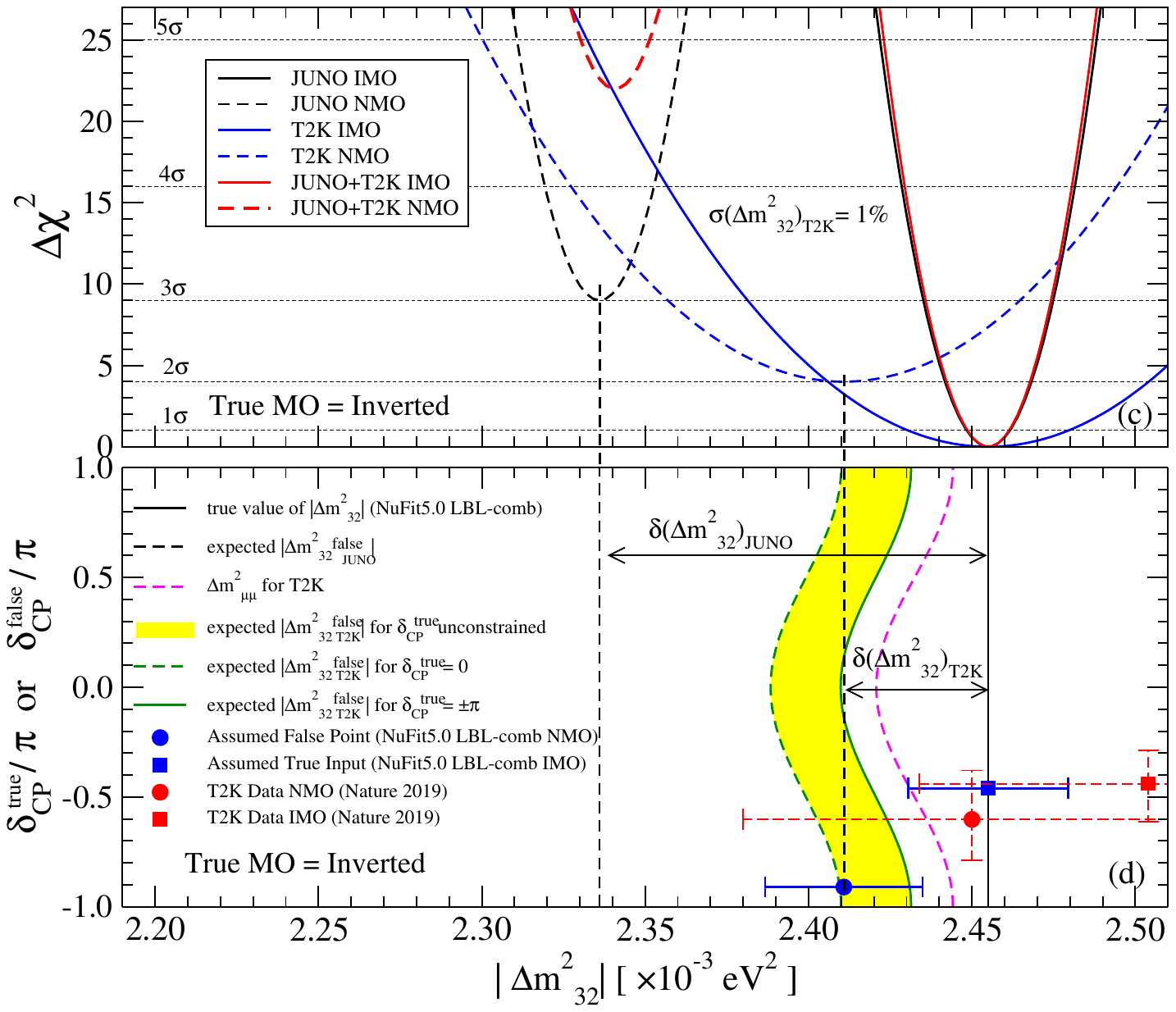}
	\vskip -0.4cm
	\caption
	{ \small 
	{\bf JUNO \& \Bs\ Mass Ordering Synergy.}
	The behaviours of \DCS\ terms (parabolas) are shown as a function of \aDm\ for JUNO (black), T2K (or \N), and their enhanced combination (red).
	The $\Delta {m^2_{32}}^\text{true}$ is fixed to the NuFit5.0 best value shown, respectively, in panels (a) and (c) for NMO and IMO.
	The extra gain in \DCSB\ discrimination numerically originates from the fact that the true \aDm\ solutions should match between JUNO (solid black vertical line) and \Bs\ (solid blue vertical circle); hence the false solutions (dashed vertical lines) must differ.
	Panels (b) and (d) illustrate this origin.
	The relation between true-false \Dm\ solutions is different and complementarity for JUNO and LB$\nu$B experiments.
	The difference is large ($\approx$ 1.5$\times$\dmN) for JUNO.
	Instead, \Bs\ exhibits a smaller difference that modulates with \dCP.
	So, the relative difference between 
	$\Delta {m^2_{32}}^\text{false}_\text{\tiny\ JUNO}$ 
	and 
	$\Delta {m^2_{32}}^\text{false}_\text{\tiny\ LB$\nu$B}$
	is maximal (minimal) for the \dCP-conserving $\pm\pi$ (0) value.
	Hence, \DCSB\ depends on \dCP\ , and ambiguity arises (yellow band) from the a priori different values of \dCP\ for the true or false solutions.
	The T2K data (red points) contrasts the precision on \aDm\ now~\cite{Abe:2019vii} compared to needed scenarios $\leq$1.0\% scenario (blue points and parabolas).
	The precision of each contribution is indicated by the parabolas' width, where JUNO is fixed to the nominal value~\cite{An:2015jdp}.
	}
\label{FigA4}
\end{figure*}

The extra synergy for MO determination sensitivity by combining JUNO and LB$\nu$B DC can be achieved thanks to the {\it mismatch} (or disagreement) of the fitted $\Delta m^2_{32}$ values for the wrong MO solutions between these two types of experiments.
For the correct MO, $\Delta m^2_{32}$ values measured by different experiments should agree with each other within the experimental uncertainties. But for those values which correspond to the wrong MO do not agree. The difference  can be quantified and used to enhance the sensitivity. 

Following the procedure described in \cite{Li:2013zyd,An:2015jdp}, we include to the JUNO analysis the external information on the external information on $\Delta m^2_{32}$ from \Bs\ with an additional pull term as 
\begin{eqnarray}
\chi^2 = {\chi^2}_\text{\tiny JUNO}+
\left(\frac{\Delta m^2_{32}-\Delta{m^2_{32}}^\text{\tiny NMO or IMO}_\text{\tiny LB$\nu$B}}
{\sigma (\Delta m^2_{32})_\text{\tiny LB$\nu$B}}
\right)^2,
\label{eq:chi2_juno_plus_pull}
\end{eqnarray}
where $\chi^2_\text{\tiny JUNO}$ implies the $\chi^2$ function 
for JUNO alone computed in a similar fashion as in \cite{An:2015jdp},
$\sigma (\Delta m^2_{32})_\text{\tiny LB$\nu$B}$ is 
the experimental uncertainty on $\Delta m^2_{32}$ 
achieved by LB$\nu$B based experiments. 
As typical values in this paper, we consider 3 cases 
$\sigma (\Delta m^2_{32})_\text{\tiny LB$\nu$B}$ = 1, 0.75 and 0.5\%.

In order to take into account the possible fluctuation of 
the central values of the measured 
$\Delta {m^2_{32}}_\text{\tiny LB$\nu$B}$
we define the extra boosting $\Delta \chi^2$ due to the synergy of 
JUNO and LB$\nu$B based experiments as 
the difference of $\chi^2$ defined in Eq.~\eqref{eq:chi2_juno_plus_pull}
for normal and inverted MO as, 
\begin{eqnarray}
\Delta \chi^2_\text{\tiny boost} \equiv 
\pm \left( \chi^2_\text{IMO}-\chi^2_\text{NMO}\right),
\label{eq:Delta_chi2_boost}
\end{eqnarray}
where +(-) sign corresponds to the case where 
the true MO is normal (inverted). 
Note that in our simplified phenomenological approach
(based on the future simulated JUNO data), for the
case with no fluctuation, 
by construction, 
$\chi^2_\text{NMO (IMO)} = 0$ for NMO (IMO).

Let us try to see how the boosting will be realized by applying our discussion to JUNO and T2K for illustration. 
In Figure ~\ref{FigA4} for the cases where the true MO is normal in the panel (a) and inverted in the panel (c) we show by the solid (dashed) black curve $\Delta \chi^2$ for JUNO alone case for true (false) MO.
The difference of $\Delta \chi^2$ between true and false MO is 9 if only JUNO is considered implying that the false MO (indicated by the dashed curves) can be rejected at 3$\sigma$.
On the other hand, let us assume the case where T2K can determine $|\Delta m^2_{32}|$ with 1\% uncertainty, and the corresponding $\Delta \chi^2$ curves are given by the solid (dashed) blue curves for true (false) MO in the same plots, rejecting the wrong MO only at 2$\sigma$ by T2K alone.
If we combine JUNO and T2K following the procedure described in this section, the resulting $\Delta \chi^2$ are given by the solid (dashed) red curves for true (false) MO, rejecting the wrong MO with more than 4$\sigma$ for both NMO and IMO. 

The large ($\sim 10$) increase of the combined $\Delta \chi^2$ for the wrong MO fit comes from the mismatch of the false $|\Delta m^2_{32}|$ values between JUNO (black dashed line) and T2K (yellow colour bands) shown in the panels (b) and (d) of Figure~\ref{FigA4}.
This is nothing the boosting effect, which can be analytically understood and quantified as follows.

\begin{figure*}[b!]
	\centering
	\vspace{-0.8cm}
	\includegraphics[scale=0.45]{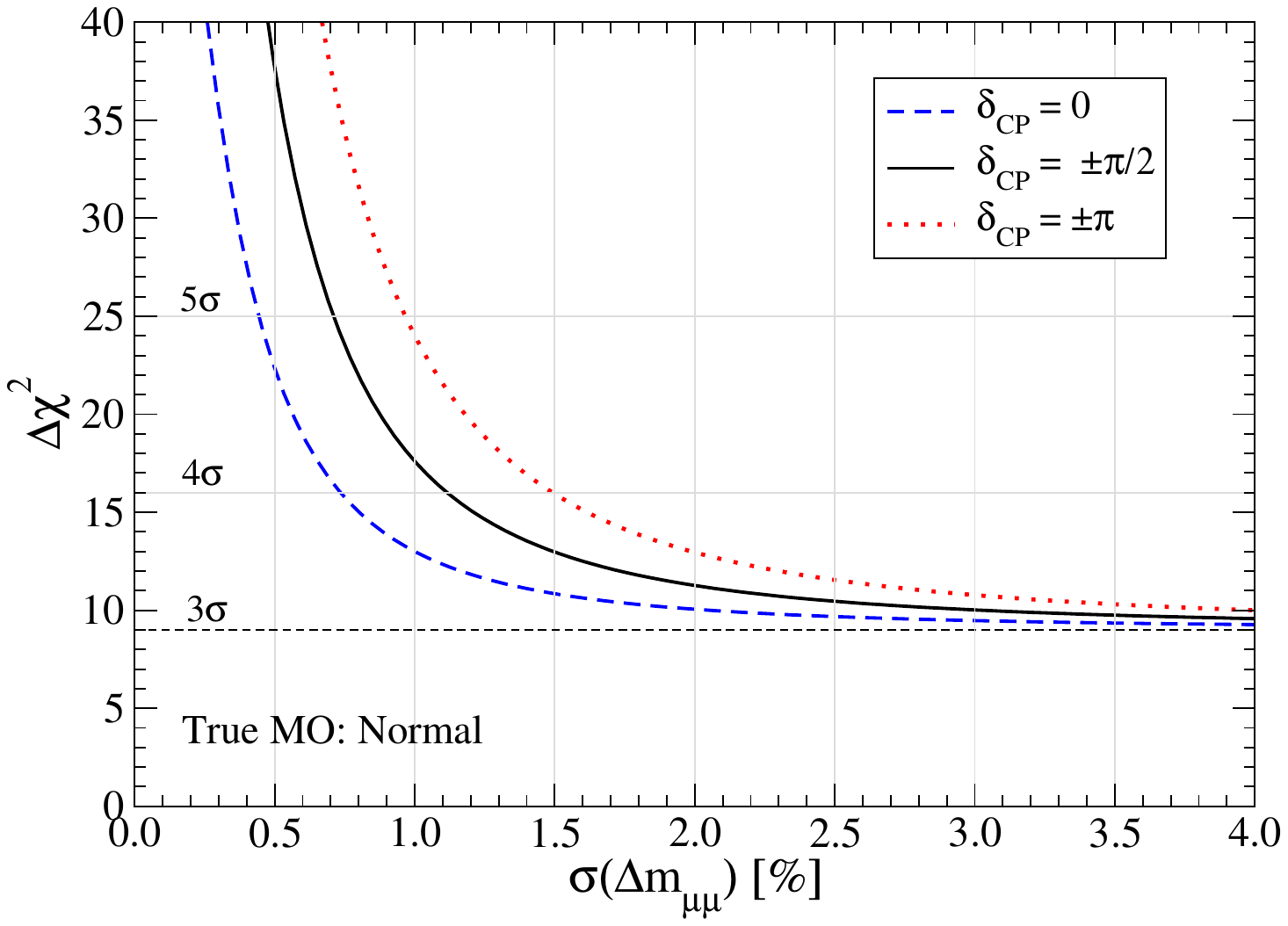}
	\vspace{-0.2cm}
	\caption
	{ \small {\bf $\mathbf{\Delta}$\CS(JUNO$\oplus$\Bs-\DCs) as a Function of the Precision of $\mathbf{\Delta m^2_{\mu\mu}}$.}
	Expected MO sensitivity to be obtained by JUNO with external information of $\Delta m^2_{\mu \mu}$ coming from LB$\nu$B experiments following the procedure described in ~\cite{Li:2013zyd,An:2015jdp}, are shown as a function of the precision of $\Delta m^2_{\mu\mu}$ for $\cos\delta_\text{\tiny CP} = \pm1$ and 0. 
	This plot is similar to Figure~7 in~\cite{Li:2013zyd}, once upgraded to the latest global data inputs~\cite{NuFit5.0}.
	We observe that they are consistent with each other, if the curves for the $\delta_\text{\tiny CP}$ values of $0^\circ$ (blue) and 180$^\circ$ (red) were interchanged, as a result of a typo in the legend of~\cite{Li:2013zyd}.
}
\label{FigA5}
\end{figure*}

Suppose that we try to perform a $\chi^2$ fit assuming the wrong MO. 
Let us first assume that $\sigma(\Delta m^2_{32})_\text{JUNO}\ll 
\sigma(\Delta m^2_{32})_\text{\tiny LB$\nu$B}$ 
and no fluctuation for simplicity
(i.e. ${\chi^2}_\text{\tiny true MO}$ =0). 
The first term in Eq.~\eqref{eq:chi2_juno_plus_pull},  
${\chi^2}_\text{\tiny JUNO}$, 
forces to drive the fitted value of $\Delta m^2_{32}$ very close to the {\it false} one favoured by JUNO or 
$\Delta {m^2_{32}}^\text{\tiny false}_\text{\tiny JUNO}$ (otherwise, ${\chi^2}_\text{\tiny JUNO}$ value increases significantly).
Then the extra increase of $\chi^2$ is approximately given by the second term in Eq.~\eqref{eq:chi2_juno_plus_pull} with $\Delta m^2_{32}$ replaced 
by $\Delta {m^2_{32}}^\text{\tiny false}_\text{\tiny JUNO}$, 
\begin{eqnarray}
\Delta \chi^2_\text{\tiny boost} 
&\sim  & 
\left[
\frac{\Delta {m^2_{32}}^\text{\tiny false}_\text{\tiny JUNO}-
\Delta{m^2_{32}}^\text{\tiny false}_\text{\tiny LB$\nu$B}}
{\sigma (\Delta m^2_{32})_\text{\tiny LB$\nu$B}}
\right]^2, \nonumber \\
& & \hskip -2.3cm \sim 
\left[
\frac{
\delta m^2_\phi+
2\delta m^2_{21}(\cos 2\theta_{12} -\sin 2\theta_{12} s_{13} \tan \theta_{23} 
\cos \delta_\text{\tiny CP}) 
}
{\sigma (\Delta m^2_{32})_\text{\tiny LB$\nu$B}}
\right]^2 
\nonumber \\
& & \hskip -2.3cm
\sim 4, 9, 16\ \text{, respectively, for}\ \delta_\text{\tiny CP} = 0, \pm \pi/2,
\pm \pi,
\label{eq:delta_chi2_boost_approx}
\end{eqnarray}
where the numbers in the last line were estimated for $\sigma (\Delta m^2_{32})_\text{\tiny LB$\nu$B}= 1$\%.
The case where $\delta_\text{\tiny CP} = \pm \pi/2$ and 
$\Delta \chi^2_\text{\tiny boost} \sim 9$ can be directly compared with
more precise results shown in Figure~\ref{Fig4}(a), see the blue solid curve at
$\delta_\text{\tiny CP}^\text{true} = \pm \pi/2$ which gives 
$\Delta \chi^2_\text{\tiny boost}\sim 8$ which is in rough agreement. 
The expression in Eq.~\eqref{eq:delta_chi2_boost_approx} is in agreement with
the one given in Eq.~(18) of \cite{Li:2013zyd} apart from 
the term $\delta m^2_\phi$ which is not so large.

\subsection*{D.~Full 3 $\nu$ versus effective 2$\nu$ formulation}

In the previous discussions found in ~\cite{Li:2013zyd,An:2015jdp}, in order to demonstrate the boosting synergy effect between JUNO and LB$\nu$B experiments, the effective mass squared differences $\Delta m^2_{ee}$ and $\Delta m^2_{\mu\mu}$, defined respectively, in Eqs.~\eqref{eq:dmee} and \eqref{eq:dm_mumu} originally found in \cite{Nunokawa:2005nx} were used.
While we used these parameters in some intermediate steps of our computations, as described in Appendix~C, we did not use these parameters explicitly in our combined $\chi^2$ describing the extra synergy between JUNO and LB$\nu$B (DC) based experiments defined in Eq.~\eqref{eq:chi2_juno_plus_pull}, as well as in the final sensitivity plots presented in this paper. 
The main advantage of using these effective mass squared differences is that no a priori assumptions have to be made about any other parameters not accessible by JUNO, in particular, CP phase ($\delta_{\text{\tiny CP}}$), whereas by using them, one must specify explicitly the $\delta_{\text{\tiny CP}}$ value as done in Ref.~\cite{Li:2013zyd}.

In order to check the consistency between our work and previous studies, we have explicitly verified that the results do not depend on the parameters used in the analysis and in the presentation of the final results, provided that that comparisons are done properly. 
In Figure~\ref{FigA5}, we show $\Delta$\CS(JUNO$\oplus$\Bs-\DCs) computed by using explicitly $\Delta m_{\mu\mu}$ (instead of using $\Delta m^2_{32}$) in our $\chi^2$ analysis as done in~\cite{Li:2013zyd,An:2015jdp}, as a function of the precision of $\Delta m^2_{\mu\mu}$.
There is general good agreement with the result shown in Figure 7 of ~\cite{Li:2013zyd}, if $\delta_\text{\tiny CP}$ curves for $0^\circ$ and 180$^\circ$ were interchanged, as described in Figure~\ref{FigA5}.

\end{multicols*}
\end{document}